\documentclass{aa} 
\usepackage{graphicx}
\usepackage{txfonts}
\usepackage{hyperref}
\usepackage{multirow}
\usepackage[capitalise]{cleveref}
\usepackage[per-mode=symbol]{siunitx}
\usepackage{graphicx}
\usepackage{textcomp}
\usepackage[geometry]{ifsym} 
\usepackage{color}
\bibpunct{(}{)}{;}{a}{}{,}

\DeclareSIUnit{\astronomicalunit}{au}

\definecolor{RED}{RGB}{255,0,0}

\newcommand{\md}{\text{d}}

\begin{document}

\title{The UK Met Office GCM with a sophisticated radiation scheme applied to the hot Jupiter HD~209458b}

\titlerunning{The UK Met Office GCM applied to HD~209458b}

\author{%
David~S.~Amundsen\inst{1,2,3}\and%
Nathan~J.~Mayne\inst{1}\and%
Isabelle~Baraffe\inst{1,4}\and%
James~Manners\inst{1,5}\and%
Pascal~Tremblin\inst{1,6}\and%
Benjamin~Drummond\inst{1}\and%
Chris Smith\inst{1,5}\and%
David~M.~Acreman\inst{1}\and%
Derek~Homeier\inst{7,4}
}
\authorrunning{Amundsen  et al.}

\institute{%
Astrophysics Group, University of Exeter, Exeter, EX4 4QL, United Kingdom
\and
Department of Applied Physics and Applied Mathematics, Columbia University, New York, NY 10025, USA\\
\email{d.s.amundsen@columbia.edu}
\and
NASA Goddard Institute for Space Studies, New York, NY 10025, USA
\and
Univ Lyon, ENS de Lyon, Univ Lyon 1, CNRS, CRAL, UMR5574, F-69007, Lyon, France
\and
Met Office, Exeter, EX1 3PB, United Kingdom
\and
Maison de la Simulation, CEA-CNRS-INRIA-UPS-UVSQ, USR 3441, Centre d'\'{e}tude de Saclay, F-91191 Gif-Sur-Yvette, France
\and
Zentrum f\"{u}r Astronomie der Universit\"{a}t Heidelberg, Landessternwarte K\"{o}nigstuhl 12, D-69117 Heidelberg, Germany
}

\date{}

\abstract{
To study the complexity of hot Jupiter atmospheres revealed by observations of increasing quality, we have adapted the UK Met Office global circulation model (GCM), the Unified Model (UM), to these exoplanets. The UM solves the full 3D Navier--Stokes equations with a height-varying gravity, avoiding the simplifications used in most GCMs currently applied to exoplanets. In this work we present the coupling of the UM dynamical core to an accurate radiation scheme based on the two-stream approximation and correlated-$k$ method with state-of-the-art opacities from ExoMol. Our first application of this model is devoted to the extensively studied hot Jupiter HD 209458b. We have derived synthetic emission spectra and phase curves, and compare them to both previous models also based on state-of-the-art radiative transfer, and to observations. We find a reasonable a agreement between observations and both our day side emission and hot spot offset, however, our night side emission is too large. Overall our results are qualitatively similar to those found by \citet{Showman2009} with the SPARC/MITgcm, however, we note several quantitative differences: Our simulations show significant variation in the position of the hottest part of the atmosphere with pressure, as expected from simple timescale arguments, and in contrast to the ``vertical coherency'' found by \citet{Showman2009}. We also see significant quantitative differences in calculated synthetic observations. Our comparisons strengthen the need for detailed intercomparisons of dynamical cores, radiation schemes and post-processing tools to understand these differences. This effort is necessary in order to make robust conclusions about these atmospheres based on GCM results.
}



\newlength{\figurewidth}
\setlength{\figurewidth}{\columnwidth}


\maketitle

\section{Introduction} \label{sec:introduction}

Hot Jupiters, Jupiter-sized planets orbiting close to their parent stars, have the most observationally constrained atmospheres of all exoplanets. Transmission spectroscopy has been used to detect sodium, potassium and water~\citep{Charbonneau2002,Snellen2008,Redfield2008,Sing2011,Sing2012,Sing2015,Deming2013,Wakeford2013,McCullough2014,Evans2016} as well as identifying a continuum of hot Jupiter atmospheres ranging from ``cloudy'' to ``clear''~\citep{Sing2016}. Temperature contrasts and brightness temperature
maps have been derived from phase curves~\citep{Knutson2007a,Knutson2009,Knutson2012,Maxted2013,Zellem2014,Stevenson2014}, and finally wind velocities have been estimated~\citep{Snellen2010,Louden2015}. These measurements have revealed that many hot Jupiters are subject to significant heat redistribution between their day and night sides, with hot spots shifted eastward of the substellar point, which can not be modelled consistently with one-dimensional (1D) models.

This has motivated the adaptation of global circulation models (GCMs), which to date have been applied to study the atmospheric circulation of several hot Jupiters~\citep{Showman2002,Cooper2005,Showman2009,Lewis2010,Rauscher2010,Thrastarson2010,Polichtchouk2012,Dobbs-Dixon2013,Kataria2013,Kataria2015,Kataria2016}. More recently, GCMs have also been used to study some aspects of cloud formation and evolution in hot Jupiter atmospheres \citep{Parmentier2016,Helling2016,Lee2016}. GCMs are three-dimensional (3D) models and include a dynamical core solving the equations of motion for the fluid, combined with a radiation scheme for treating stellar heating and thermal cooling of the atmosphere. Many GCMs applied to hot Jupiters solve the primitive equations~\citep[see e.g.][]{Showman2002,Showman2009,Rauscher2010,Kataria2013,Kataria2015}, which are an approximation to the full Navier--Stokes equations assuming that the atmosphere is shallow compared to the radius of the planet, in hydrostatic equilibrium and has a gravity constant with height. The exceptions are \citet{Dobbs-Dixon2013} and \citet{Mayne2014a}, who solved the full Navier--Stokes equations. As the vertical extent of hot Jupiter atmospheres can be about $\SI{10}{\percent}$ of the planet radius, the validity of the primitive equations is questionable~\citep{Mayne2014a}.

Radiation schemes in initial hot Jupiter GCMs employed Newtonian forcing, where the temperature is relaxed linearly towards equilibrium $P$--$T$ profiles~\citep{Showman2002,Cooper2005,Cooper2006,Showman2008,Rauscher2010}. Such approaches have many disadvantages as radiative heating and cooling are not treated self-consistently: (i) appropriate equilibrium profiles are difficult to obtain from 1D models, (ii) the temperature relaxation is linear while in reality it may be non-linear for large deviations from the equilibrium profiles, (iii) atmospheric interactions due to exchange of radiative energy such as emission and absorption of thermal radiation are ignored and (iv) the model flexibility is poor since for each new planet modelled, the forcing must be changed.

More recent hot Jupiter GCMs have adopted radiation schemes using the two-stream approximation with grey~\citep{Rauscher2012} or average opacity schemes~\citep{Dobbs-Dixon2013}, which has recently been shown to yield inaccurate heating rates when considering molecular absorption in these atmospheres~\citep{Amundsen2014}. The most sophisticated radiation scheme employed to date is that presented in \citet{Showman2009} and later used in \citet{Lewis2010} and \citet{Kataria2013,Kataria2015,Kataria2016}, which adopts the two-stream approximation~\citep{Thomas2002} for the stellar component and the two-stream source function technique~\citep{Toon1989} for the thermal component, combined with the correlated-$k$ method~\citep{Lacis1991} for treating opacities. This has been shown to yield significantly better agreement with observations compared to using Newtonian forcing~\citep{Showman2008,Showman2009}. \citet{Amundsen2014} showed that the two-stream approximation and correlated-$k$ method give accurate fluxes and heating rates for hot Jupiter atmospheres.

We have adapted the UK Met Office GCM, the Unified Model (UM), for the study of hot Jupiters. The UM dynamical core solves the full 3D Navier--Stokes equations with a height-varying gravity, and the radiation scheme is state-of-the-art using the two-stream approximation and correlated-$k$ method to treat opacities. The adaptation of the dynamical core~\citep{Mayne2014a,Mayne2014b} and radiation scheme~\citep{Amundsen2014} have been presented in previous publications. Preliminary results from our model have been presented in \citet{Helling2016}, however, here we use an updated opacity database and present the model and results in more detail.

The goal of the present work is to (i) provide the technical details of the coupling between the UM dynamical core and adapted radiation scheme for hot Jupiters for future reference, (ii) provide the first comparison between two hot Jupiter GCMs \citep[ours and that of][]{Showman2009} with similar state-of-the-art radiation schemes and investigate the robustness of these GCMs, (iii) evaluate differences in resulting synthetic observations calculated from model output as the model of \citet{Showman2009} has already been used extensively in the literature~\citep[see e.g.][]{Agundez2014,Fortney2010,Kataria2015,Moses2011,Wong2015,Zellem2014}, and (iv) provide a guide for future more in-depth intercomparisons of these models.

To ease comparison with earlier GCMs and the models of \citet{Showman2009}, we use parameters similar to those of HD~209458b. This is the first attempt to compare results obtained with two different hot Jupiter GCMs with sophisticated radiation schemes. Both dynamical cores\footnote{See \url{http://earthsystemcog.org/projects/dcmip-2012/}.} and radiation schemes \citep{Ellingson1991,Collins2006,Oreopoulos2012} are tested thoroughly through intercomparison projects for Earth-like conditions. Intercomparison of hot Jupiter GCMs will in the future become crucially important as the quality of observations improve.

In contrast to \citet{Showman2009}, we do not include TiO and VO in our model. Unfortunately the model becomes unstable due to the intense heating of the upper atmosphere caused by these molecules, a consequence of their large opacity at visible wavelengths. This prevents a detailed comparison to the model of HD~209458b in \citet{Showman2009}, however, at present there is no evidence for TiO and VO, or even a temperature inversion, in the atmosphere of this planet~\citep{Diamond-Lowe2014,Hoeijmakers2014,Evans2015,Schwarz2015,Line2016}.

This paper is organised as follows: In \cref{sec:model_description} we briefly describe the model, including the dynamical core and radiation scheme. In \cref{sec:forcing} we discuss results from running the model of \citet{Mayne2014a} which uses Newtonian forcing in place of an accurate radiation scheme before discussing results from our coupled model with sophisticated radiative transfer in \cref{sec:coupled_um_ntv}. Synthetic observations are calculated for all models presented and compared to available observations of HD~209458b in the literature. Our conclusions are presented in \cref{sec:conclusions}.

\section{Model description} \label{sec:model_description}

In this section we briefly describe the dynamical core (\cref{sec:dynamics}) and radiation scheme (\cref{sec:radiation_scheme}), but refer to \citet{Mayne2014a}, \citet{Amundsen2014}, \citet{Amundsen2015thesis} and \citet{Amundsen2016b} for more details.

\subsection{Dynamics} \label{sec:dynamics}

We used the Met Office Unified Model (UM) with the ENDGAME (Even Newer Dynamics for General Atmospheric Modelling of the Environment) dynamical core~\citep{Wood2014} to solve the non-hydrostatic, deep-atmosphere Navier--Stokes equations for planetary atmospheres with a height-varying gravity. The equations are solved on a latitude--longitude--height grid using a semi-Lagrangian semi-implicit scheme. We use free-slip impermeable upper and lower boundaries located at a fixed height. We have previously applied ENDGAME to HD~209458b successfully using a Newtonian forcing scheme~\citep{Mayne2014a}, and we refer to this paper and references therein for more details on the dynamical core. Here we exclusively solve the ``full'' equation set solving the non-hydrostatic deep-atmosphere equations with a height-varying gravity.


The diffusion scheme is described in \citet{Mayne2014a,Mayne2014b}, and includes separate components in the longitude ($K_{\lambda}$) and latitude ($K_{\phi}$) directions. The UM numerical scheme does not explicitly enforce axial angular momentum conservation (AAM), however, in the results presented here AAM is conserved to better than $\SI{98}{\percent}$. AAM conservation was maximised by using $K_{\lambda} \sim 0.16$ and $K_{\phi} = 0.0$, which is applied in all simulations presented in this work and those presented in \citet{Mayne2014a}, where the latter incorrectly reported the values of these constants. As described in \citet{Mayne2014a,Mayne2014b}, due to the particular horizontal grid staggering adopted at the pole, we do not need to use a polar filter, but note that the diffusion scheme has some aspects in common with a polar filter.

\subsection{Radiation scheme} \label{sec:radiation_scheme}

To calculate radiative heating rates we used the Suite of Community Radiative Transfer codes based on Edwards and Slingo (SOCRATES) scheme\footnote{\url{https://code.metoffice.gov.uk/trac/socrates}}~\citep{Edwards1996a,Edwards1996b}, which uses the two-stream approximation combined with the correlated-$k$ method for both the stellar and thermal components. We have presented the adaptation and testing of this radiation scheme for hot Jupiter-like atmospheres in \citet{Amundsen2014}, where it was found to yield accurate fluxes and heating rates by comparing to discrete ordinate line-by-line calculations. Here we review modifications made to the radiation scheme since it was presented in \citet{Amundsen2014}.

\subsubsection{The radiative transfer equation}

For the thermal component of the radiation we solve the two-stream equations as formulated by \citet{Zdunkowski1985} and \citet{Edwards1996b} with no scattering and a diffusivity $D = 1.66$, which was found to yield the most accurate fluxes and heating rates in \citet{Amundsen2014}. For the stellar component we solve the two-stream equations as formulated by \citet{Zdunkowski1980}, which uses $D = 2$. Rayleigh scattering by H$_2$ and He is included, with refractive indices for H$_2$ and He from \citet{Leonard1974} and \citet{Mansfield1969}, respectively\footnote{Data collected from \url{http://refractiveindex.info/}.}, and an anisotropy factor $\rho_n = 0.02$~\citep{Penndorf1957}. Refractive indices for H$_2$ and He are combined using the Lorentz--Lorentz relation~\citep{Heller1965}.

Vertically, the dynamical core defines potential temperatures in layers and exner pressures on levels (layer interfaces). The radiation scheme uses the same grid, with fluxes calculated at the levels defined by the exner pressure, and layer properties are set based on the potential temperature. As temperatures are also required at the exner pressure levels by the radiation scheme they are interpolated linearly in height from the layer values. Layer radiative heating rates can then be calculated and applied directly by differencing fluxes at neighbouring levels without interpolation.

\subsubsection{Opacities} \label{sec:opacities}

Our opacity database includes absorption by H$_2$O, CO, CH$_4$, NH$_3$, and H$_2$--H$_2$ and H$_2$--He collision induced absorption (CIA) as described in \citet{Amundsen2014}, using the newest ExoMol line lists~\citep{Tennyson2012,Tennyson2016} where available, with the recent addition of the alkali metals Li, Na, K, Rb and Cs. Alkali metal opacities are included using transition probabilities and broadening coefficients from VALD3~\citep{Heiter2008}. Voigt profiles are used for all lines with a line cut-off at $\SI{4000}{\centi \metre^{-1}}$ from the line centres except for the Na and K D lines for which we use line profiles from the PHOENIX atmosphere code~\citep[][Derek Homeier, priv. comm.]{Allard1999,Allard2003,Allard2007}. We have updated our CH$_4$ opacities to use the YT10to10 CH$_4$ line list~\citep{Yurchenko2014}.

Opacities are treated using the correlated-$k$ method~\citep{Lacis1991} as described in \citet{Amundsen2014}, but $k$-coefficients are here computed individually for each gas. Our $32$ bands are defined in \citet{Amundsen2014}, and in each band the main absorber is found by comparing transmissions using the maximum equilibrium abundance for each gas. $k$-coefficients for individual gases are combined on-the-fly in the UM using equivalent extinction~\citep{Edwards1996b}, where all gases except the strongest absorber in each band is taken into account through a grey absorption. The direct stellar component, however, is computed directly by multiplying transmissions for each gas, which assumes absorption lines for different gases are randomly overlapping~\citep{Lacis1991}, using all $k$-coefficients for all gases. We have found this approach to be more accurate than using a pre-computed table of $k$-coefficients for the gas mixture as used in \citet{Amundsen2014} and \citet{Showman2009} as the use of such tables involves interpolating both mixing ratios and gas opacities in temperature and pressure, not only the opacities of individual absorbers. For a more detailed discussion and a comparison of different treatments of overlapping gaseous absorption, see~\citep{Amundsen2016b}.

\subsubsection{Abundances} \label{sec:abundances}

Abundances were calculated as in \citet{Amundsen2014} using the analytical chemical equilibrium abundance formulas for H$_2$O, CO, CH$_4$ and NH$_3$ from \citet{Burrows1999}. The alkali metal abundances are approximated by assuming they are in atomic gaseous form above the chemical transformation temperature, $T_\text{trans}^i(P)$ for alkali metal $i$, and that for $ T < T_\text{trans}^i(P)$ their atomic gaseous abundance is negligible. We take the chemical transformation curves for the alkali metal chlorides from \cite{Burrows1999} and apply an additional smoothing of the form
\begin{equation}
\phi^i(T) = \frac{1}{e^{-(T - T_\text{trans}^i)/\Delta T_\text{char}^i} + 1},
\label{eq:phi_abundance}
\end{equation}
where $\phi^i(T)$ is the normalised abundance, $T_\text{trans}^i$ is the chemical transformation temperature and $\Delta T_\text{char}^i$ is the characteristic scale over which the abundance changes for species $i$. We adopt $\Delta T_\text{char}^i = \SI{20}{\kelvin}$ for alkali metals. Physically this is a primitive way of taking into account the transition between for example Na and NaCl and avoids numerical problems associated with non-continuous abundance changes. We note, however, that the particular functional form of this smoothing has no physical basis, but was chosen as it is symmetric about $T_\text{trans}^i$, and it and all its derivatives are continuous functions of temperature.

\subsubsection{Boundary conditions}

An extra layer is included in the radiation scheme to account for absorption, emission and scattering above the dynamically modelled domain. The layer extends up to zero pressure, and extensive testing has shown that the absorption is accurately taken into account with temperatures extrapolated linearly in log pressure and the temperature at zero pressure set to the smallest temperature in our $P$--$T$ grid (\SI{70}{\kelvin}). At the lower boundary we impose a net intrinsic flux $F_\text{int} = \sigma T_\text{int}^4$, thereby taking into account heat escaping from the planet interior. The thermal upward flux at the lower boundary surface, $F_\text{surf}^+ = \sigma T_\text{surf}^4$, is then given by
\begin{equation}
F_\text{surf}^+ = \sigma T_\text{int}^4 + F_\text{surf}^-.
\label{eq:F_surf^+}
\end{equation}
where $F_\text{surf}^-$ is the downward flux at the lower boundary. To ease implementation we use the value of $F_\text{surf}^-$ in \cref{eq:F_surf^+} from the previous radiation time step. Consequently, the value of $T_\text{surf}$, and therefore the upward surface flux $F_\text{surf}^+$ used in the lower boundary condition, will lag one radiative time step behind the radiative transfer calculation. We have found the temporal variations in $F_\text{surf}^+$ to be very small compared to the radiative time step, which ensures the validity of this approximation. More details on the boundary conditions are presented in \citet{Amundsen2015thesis}.

\subsection{Synthetic observations}

We have calculated synthetic observations from UM output using our 1D discrete ordinate radiation code \texttt{ATMO}~\citep{Amundsen2014,Tremblin2015,Tremblin2016,Drummond2016}. It uses the same opacity sources as described in \cref{sec:opacities}, and to compute synthetic observations we use high resolution $k$-tables with $\num{5000}$ bands with band limits evenly spaced at $\SI{10}{\centi \metre^{-1}}$ intervals. We note that the use of high resolution $k$-tables is necessary as a line-by-line approach is too computationally expensive and a reduced line-by-line resolution of $\sim \SI{1}{\centi \metre^{-1}}$ yields very large errors in band-integrated fluxes. Chemical equilibrium abundances are calculated using a Gibbs energy minimisation scheme~\citep{Drummond2016}. Our calculations of emission spectra and phase curves from UM output are detailed below.

\subsubsection{Emission spectra}

\begin{figure}
\centering
\small{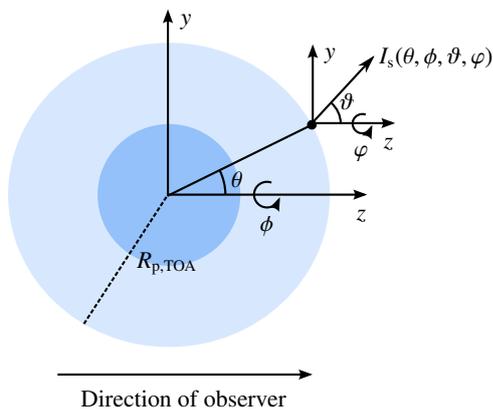}
\caption{Illustration of the definition of the angles used to calculate the hemispherically integrated emission, see \cref{eq:F_emitted}. $(\theta, \phi)$ and $(\vartheta, \varphi)$ denote the position and the direction of the radiation, respectively, with the coordinate system placed such that the $z$-axis points towards the observer.}
\label{fig:emission_schematic}
\end{figure}

The emission from a planet as measured on Earth is given by~\citep{Seager2010b}
\begin{equation}
F_\text{o} = \left( \frac{R_\text{p,TOA}}{D_\text{o}} \right)^2 \int_0^{2\pi} \int_0^{\pi/2} I_\text{s}(\theta, \phi, \vartheta_\text{o}, \varphi_\text{o}) \cos \theta \sin \theta \, \md \theta \md \phi ,
\label{eq:F_emitted}
\end{equation}
where $R_\text{p,TOA}$ is the planet radius at the top of the atmosphere, $D_\text{o}$ is the distance to the observer, and $I_\text{s}(\theta, \phi, \vartheta_\text{o}, \varphi_\text{o})$ is the intensity at the top of the atmosphere at the location defined by the polar angle $\theta$ and azimuth angle $\phi$, which can be directly related to the latitude and longitude, in the direction of the observer $(\vartheta_\text{o}, \varphi_\text{o})$, where $\vartheta_\text{o}$ is the polar angle and $\varphi_\text{o}$ is the azimuth angle. The definitions of these angles are illustrated in \cref{fig:emission_schematic}. The coordinate system in which both the location $(\theta, \phi)$ and direction $(\vartheta, \varphi)$ of the radiation are defined is placed so that the $z$-axis always points towards the observer, that is, $\vartheta_\text{o} = 0$ and the angle between the planet surface normal at $(\theta, \phi)$ and the direction of the observer is $\theta$.

The intensity $I_\text{s}$ is calculated at $16$ discrete angles determined by the Gauss-Legendre points of the discrete ordinate method for all atmospheric columns from the UM and interpolated to obtain the intensity in the direction of the observer. These intensities, $I_\text{s}(\theta, \phi, \vartheta_\text{o}, \varphi_\text{o})$, are then integrated according to \cref{eq:F_emitted} to obtain the observed emitted flux.

\subsubsection{Phase curves}

Phase curves are the emission from the planet, as viewed from Earth, as a function of time or orbital phase angle. The integrated emission as a function of orbital phase is given by \cref{eq:F_emitted} for different observer directions, which are given by the phase angle $\alpha \in [\ang{0},\ang{360})$, where $\alpha = 0$ is primary eclipse and $\alpha = \ang{180}$ is secondary eclipse. Assuming the planet is tidally locked and in a steady state the intensity at the top of the atmosphere for a given latitude and longitude will be constant as a function of time. This simplification enables us to calculate $I_\text{s}(\theta, \phi, \vartheta, \varphi)$ only once for the entire phase curve, greatly decreasing the computation time. As in \citet{Showman2009} we ignore the small inclination of the orbit~\citep{Fortney2006b}.

\subsection{Model setup and parameters}

Our model setup is similar to that used in \citet{Mayne2014a} with a few modifications. We provide our adopted model parameters in \cref{tbl:default_parameters_hd209}. Compared to \citet{Mayne2014a} the planet radius at the lower boundary of the model, $R_\text{p}$, has been decreased to account for the vertical extent of the atmosphere, the specific heat capacity, $c_P$, has been changed to be in agreement with \citet{Showman2009}, the specific gas constant, $\mathcal R = R/\bar m$, has been changed to account for the mean molecular weight $\bar m = \SI{2.3376}{\gram \per \mole}$ used in \citet{Amundsen2014}, the pressure at the lower boundary, $P_\text{bottom}$, has been chosen to be in agreement with \citet{Showman2009} and the height of the upper boundary has been slightly adjusted to account for the modified atmospheric scale height. We find that, for both numerical stability and accuracy, we need to use a dynamical time step of $\SI{30}{\second}$, much smaller than the time step used in \citet{Mayne2014a}, but in agreement with time steps used by \citet{Showman2009}.

\begin{table*}
\centering
\caption{Model parameters adopted for HD~209458b. Values are similar to those in \citet{Mayne2014a} and \citet{Showman2009}, differences are explained in the text.}
\begin{tabular}{l|r}
Parameter & Value \\ \hline
Radius, $R_\text{p}$ & $\SI{9.0e7}{\metre} = 1.259 R_\text{Jup}$ \\
Mass, $M_\text{p}$ & $\SI{1.31e+27}{\kilo \gram} = 0.690 M_\text{Jup}$ \\
Intrinsic temperature, $T_\text{int}$ & $\SI{100}{\kelvin}$ \\
Specific heat capacity, $c_P$ & $\SI{1.3e4}{\joule \per \kilo \gram \per \kelvin}$ \\
Specific gas constant, $\mathcal R = R/\bar m$ & $\SI{3556.8}{\joule \per \kilo \gram \per \kelvin}$ \\
Lower boundary pressure, $P_\text{bottom}$ & $\SI{2e7}{\pascal} = \SI{200}{\bar}$ \\
Rotation rate, $\Omega$ & $\SI{2.06e-5}{\second^{-1}}$ \\
Vertical damping coefficient & $0.15$ \\
Height of upper boundary & $\SI{9e6}{\metre}$ \\
Horizontal resolution & $144 \times 90$ \\
Vertical resolution & $66$ \\
Dynamical time step & $\SI{30}{\second}$ \\
Radiative time step & $\SI{150}{\second}$
\end{tabular}
\label{tbl:default_parameters_hd209}
\end{table*}

It is common practice in GCMs to call the radiation scheme, that is, update fluxes and heating rates, less frequently than every dynamical time step. This is done mainly for computational efficiency, and is possible as changes in fluxes and heating rates may be small between dynamical time steps. We have tried several different radiative time steps, from calling the radiation scheme every dynamical time step to calling it every ten dynamical time steps, and have found that calling it every five dynamical time steps is a good compromise between numerical accuracy and computational cost. This leads to a radiative time step of $\SI{150}{\second}$.

We initialise our model with a $P$--$T$ profile from our radiative-convective equilibrium code \texttt{ATMO}~\citep{Amundsen2014,Tremblin2015,Tremblin2016}. Profiles are calculated using $\mu_0 = \cos \theta_0 = 0.5$, where $\theta_0$ is the star zenith angle, which corresponds to a day-side average and reduces the stellar flux at the top of the atmosphere by a factor of $1/2$. To obtain a globally averaged $P$--$T$ profile the top of the atmosphere flux is further reduced by a factor $1/2$ to account for redistribution to the night-side. All models are initialised with zero winds.

\section{Model with Newtonian forcing from \citet{Mayne2014a}} \label{sec:forcing}

Before discussing results from the coupled model we briefly summarise the results from running the model of \citet{Mayne2014a}, which uses the Newtonian forcing scheme described in \citet{Cooper2005,Cooper2006}, \citet{Rauscher2010} and \citet{Heng2011}. This enables us to compare the UM and SPARC/MITgcm without the additional complication of the radiation schemes used. The model setup used here is identical to \citet{Mayne2014a}, with equilibrium $P$--$T$ profiles and timescales are from \citet{Iro2005}. We initialise the model using an average between the day and night side $P$--$T$ profiles with zero winds. In \citet{Mayne2014a} results are averaged temporally from \SIrange{200}{1200}{\day} (\si{\day} denotes Earth days) as prescribed by the \citet{Heng2011} benchmarks. We have run the model for $> \SI{1000}{\day}$, and, in contrast to \citet{Mayne2014a} who computed temporal averages from \SIrange{200}{1200}{\day}, we generally show model results after \SI{1600}{\day} with no temporal averaging as is most common in studies applying GCMs to hot Jupiters~\citep{Showman2008,Showman2009,Kataria2013,Kataria2015}. We observe temperatures to reach an approximate steady-state after \SI{1000}{\day} for $P < \SI{e5}{\pascal}$, which is expected to be the observable part of the atmosphere (see \cref{sec:forcing_discussion}). Longer simulation times will be needed to study the deeper $P > \SI{e5}{\pascal}$ regions. This is in agreement with previously published hot Juptier GCMs with simplified forcing~\citep{Showman2008}.

\subsection{Results} \label{sec:forcing_results}

\begin{figure*}
\centering
\includegraphics[width=0.75\figurewidth]{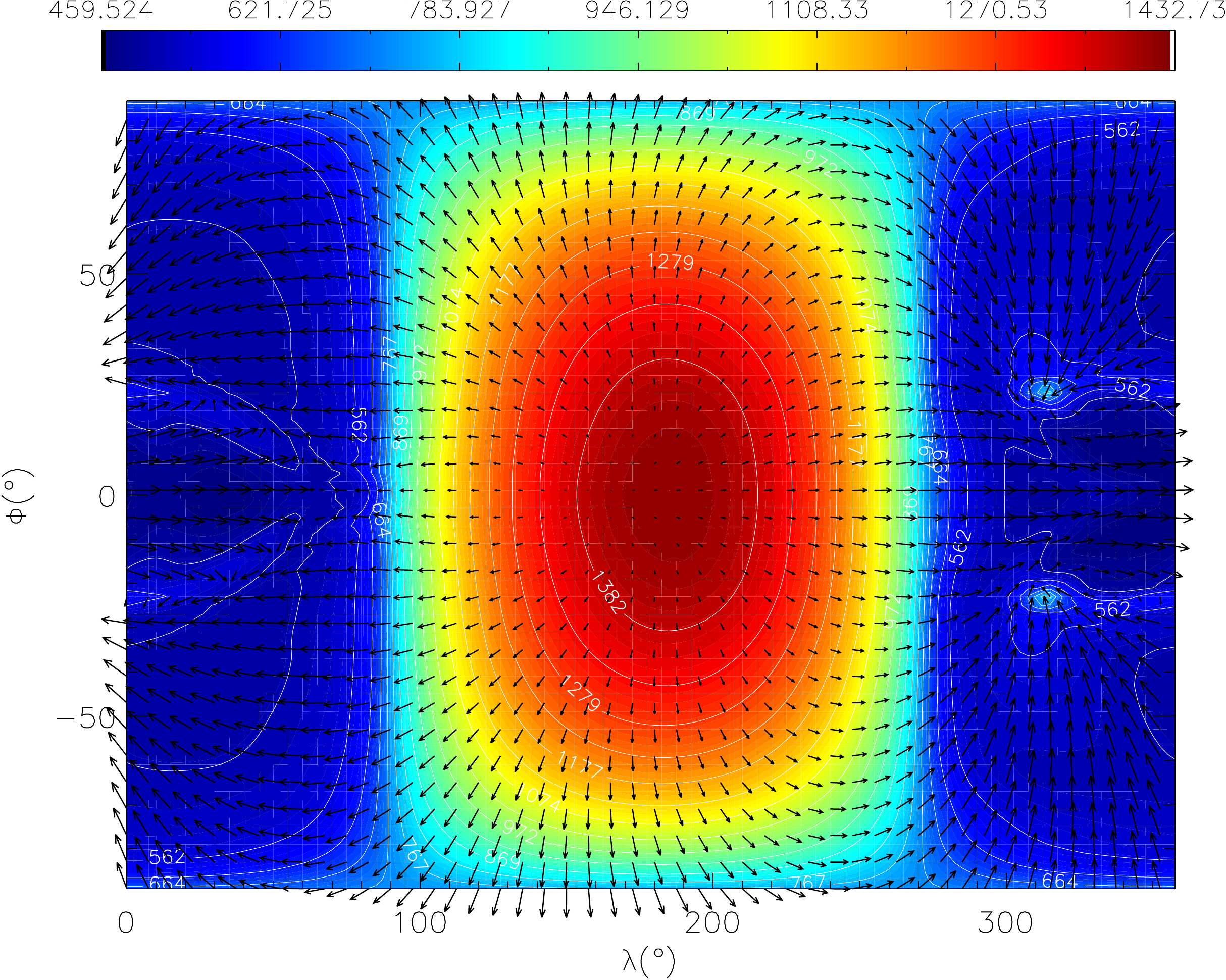}
\includegraphics[width=0.75\figurewidth]{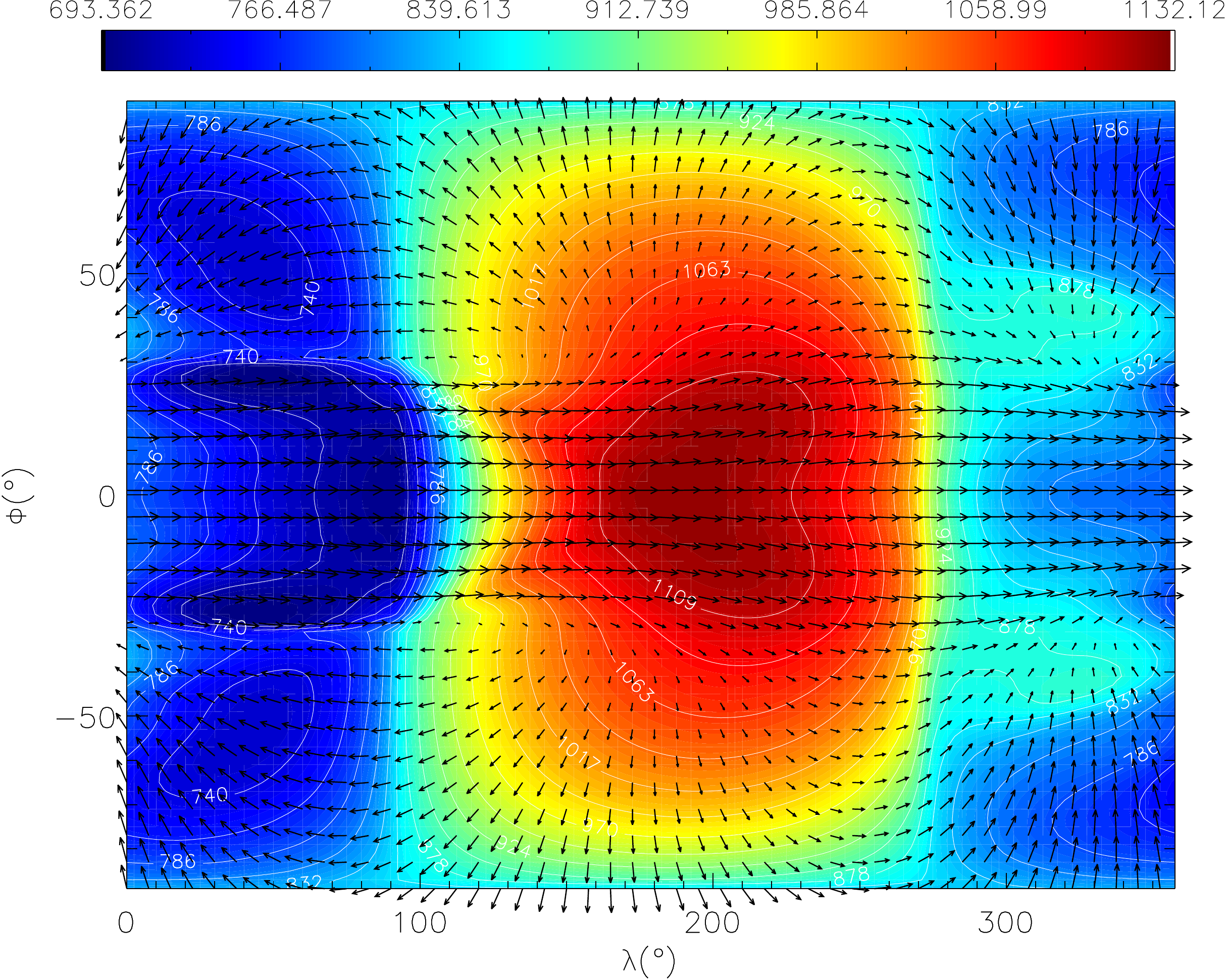}\\ \vspace{0.5cm}
\includegraphics[width=0.75\figurewidth]{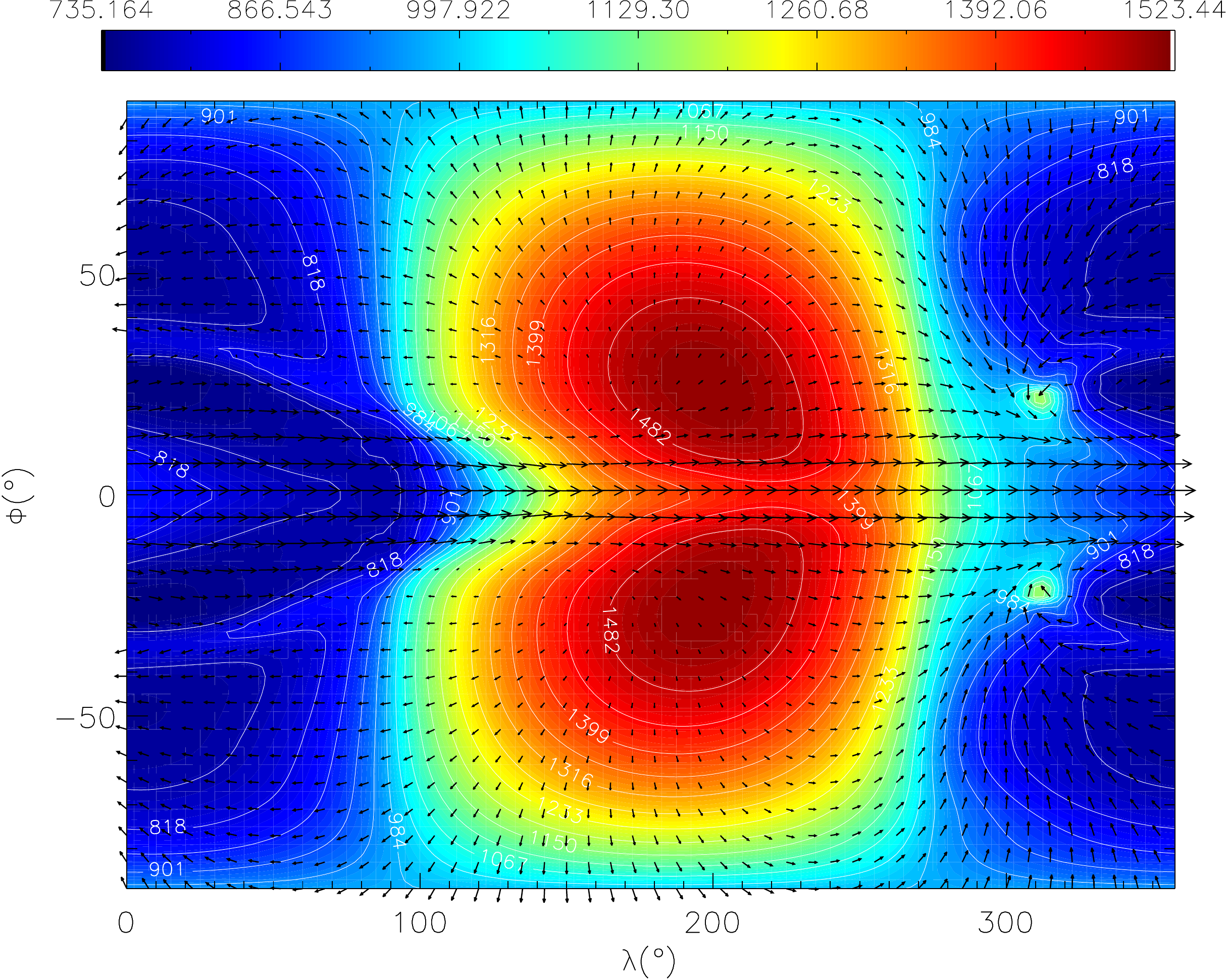}
\includegraphics[width=0.75\figurewidth]{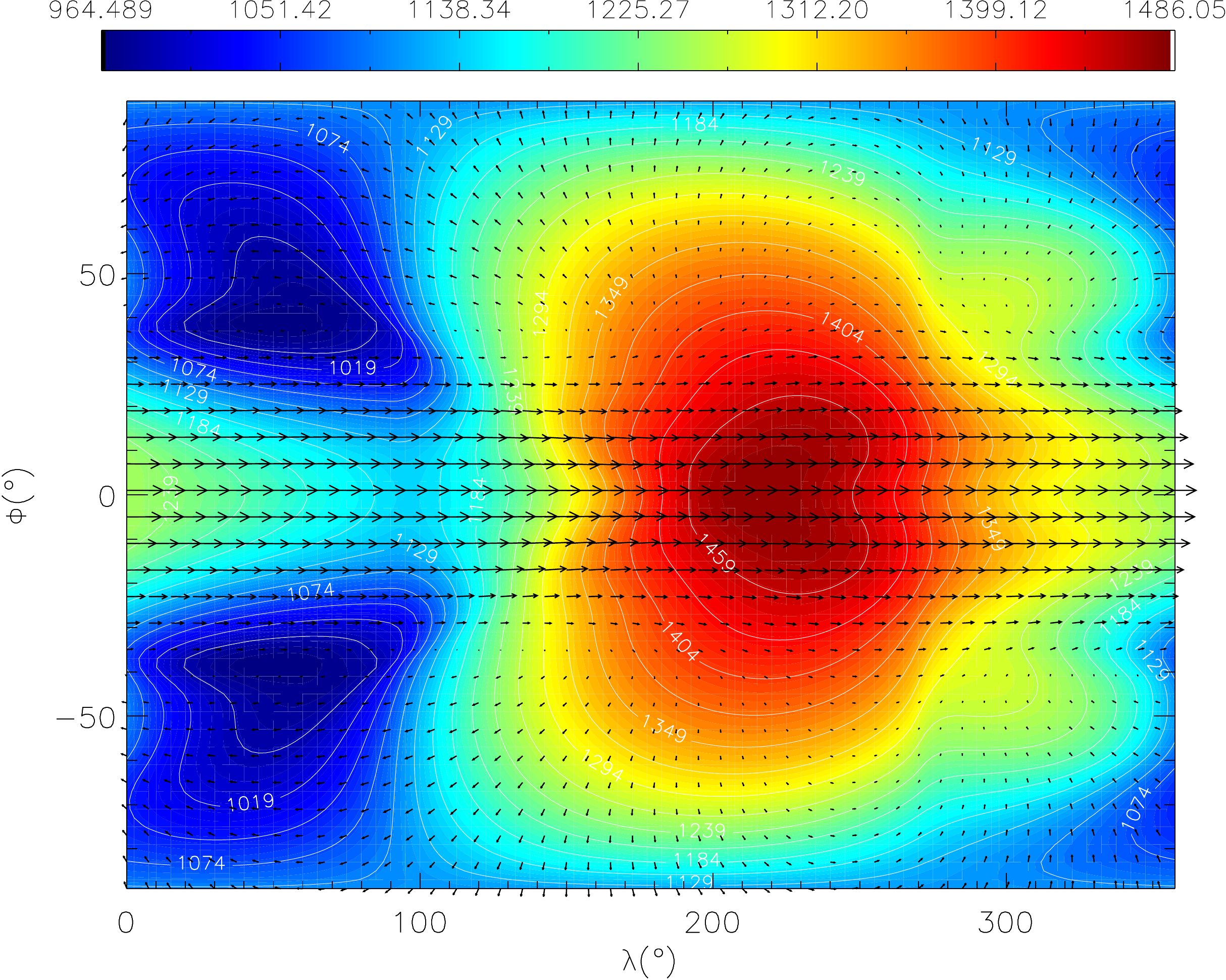}\\ \vspace{0.5cm}
\includegraphics[width=0.75\figurewidth]{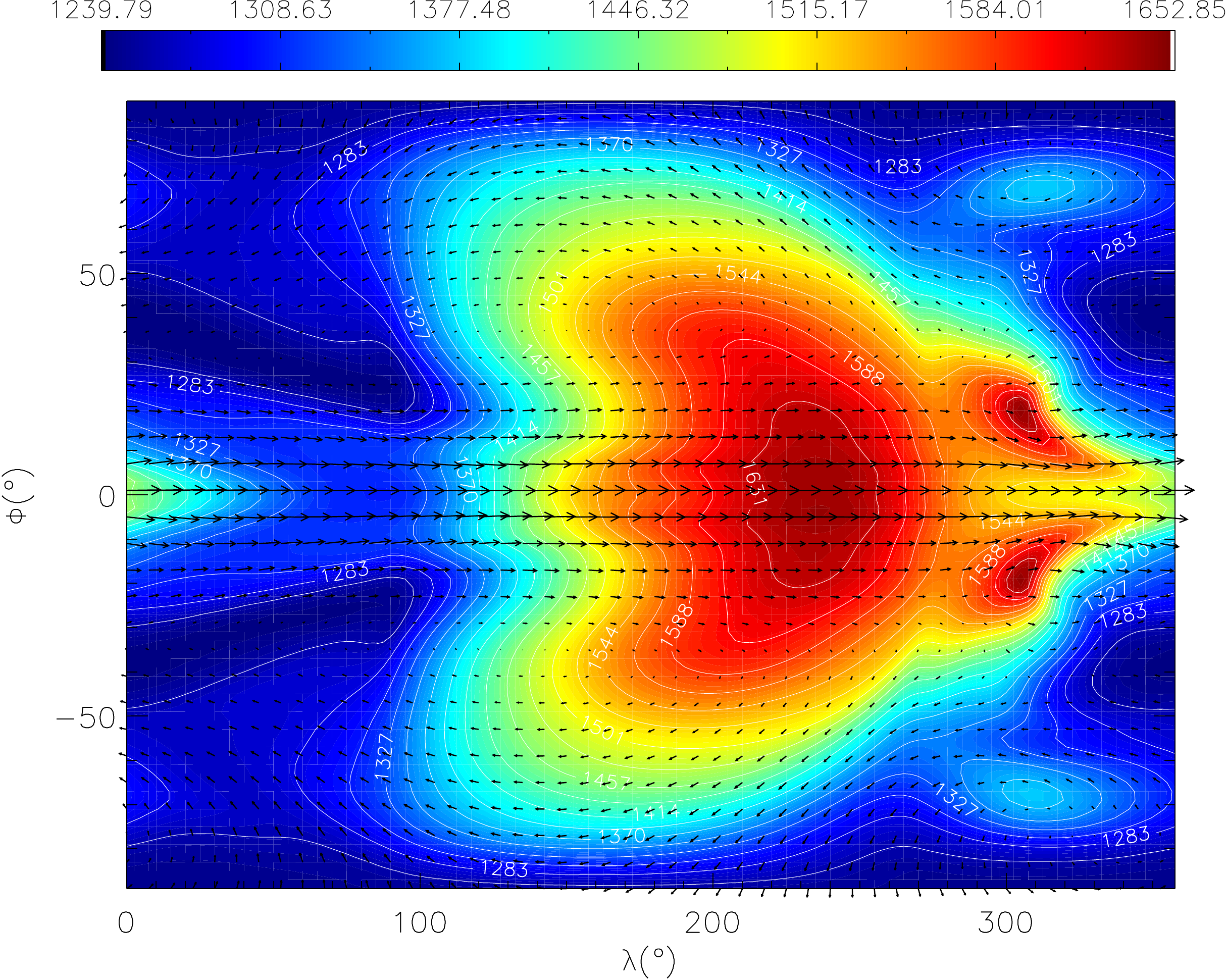}
\includegraphics[width=0.75\figurewidth]{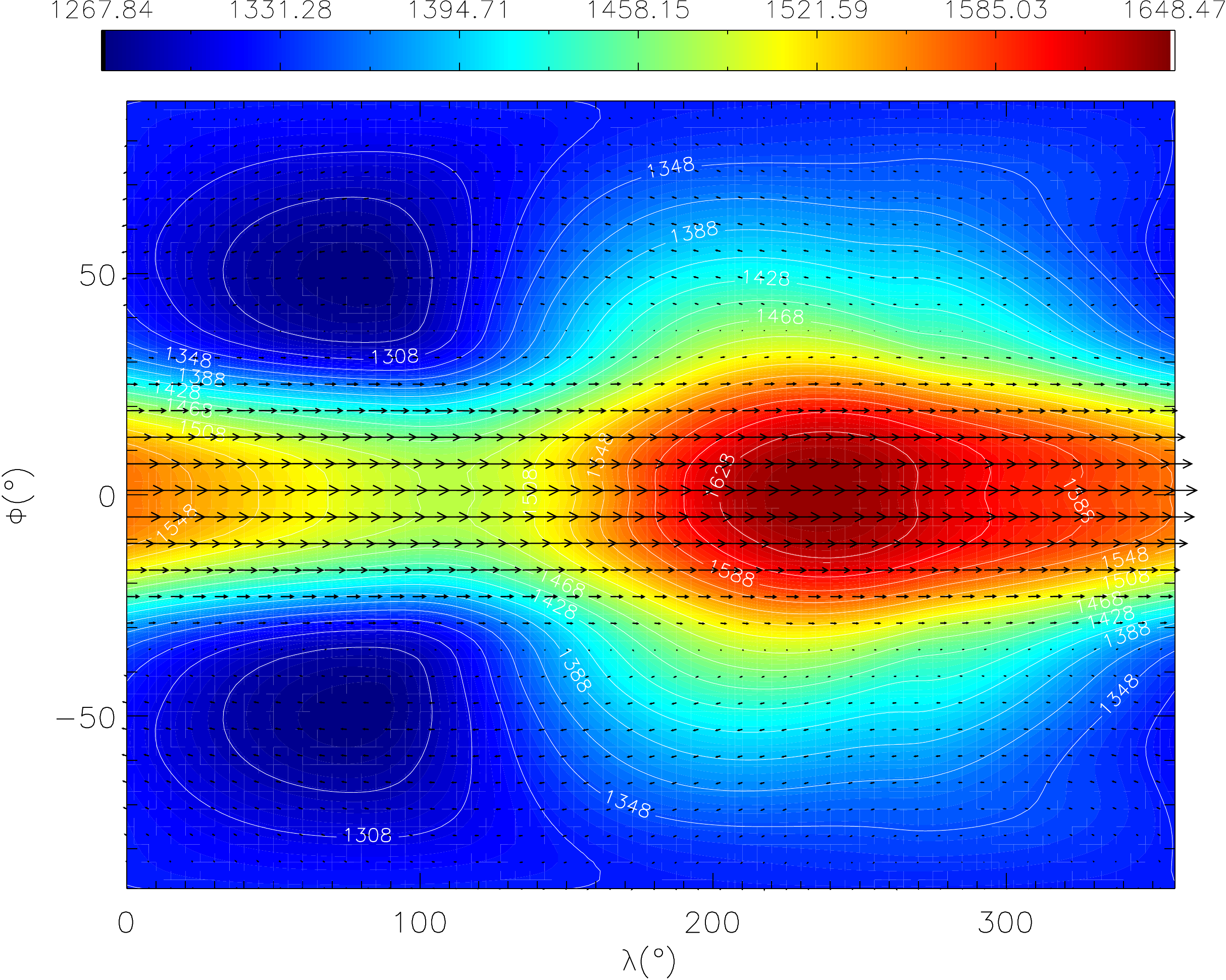}\\ \vspace{0.5cm}
\includegraphics[width=0.75\figurewidth]{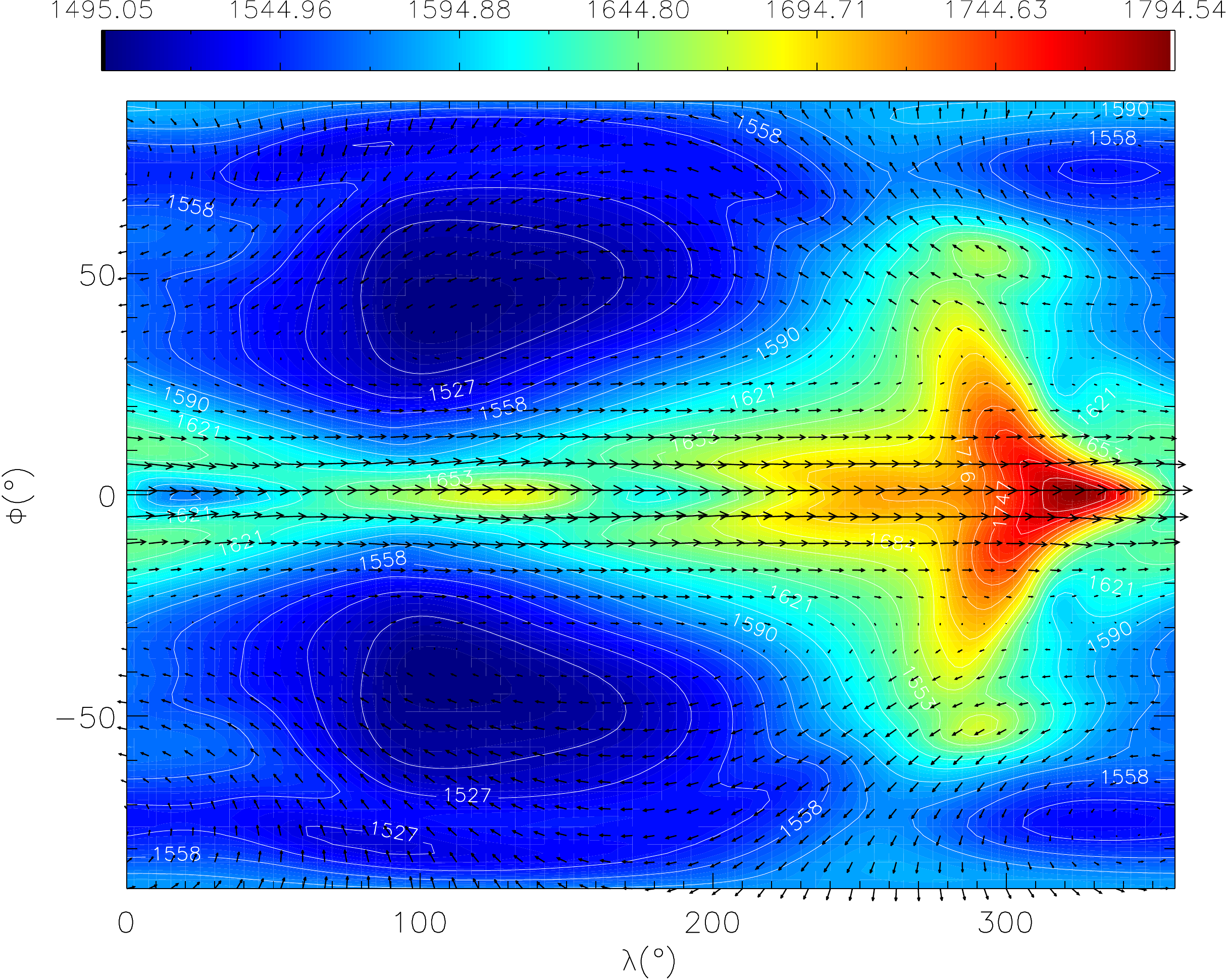}
\includegraphics[width=0.75\figurewidth]{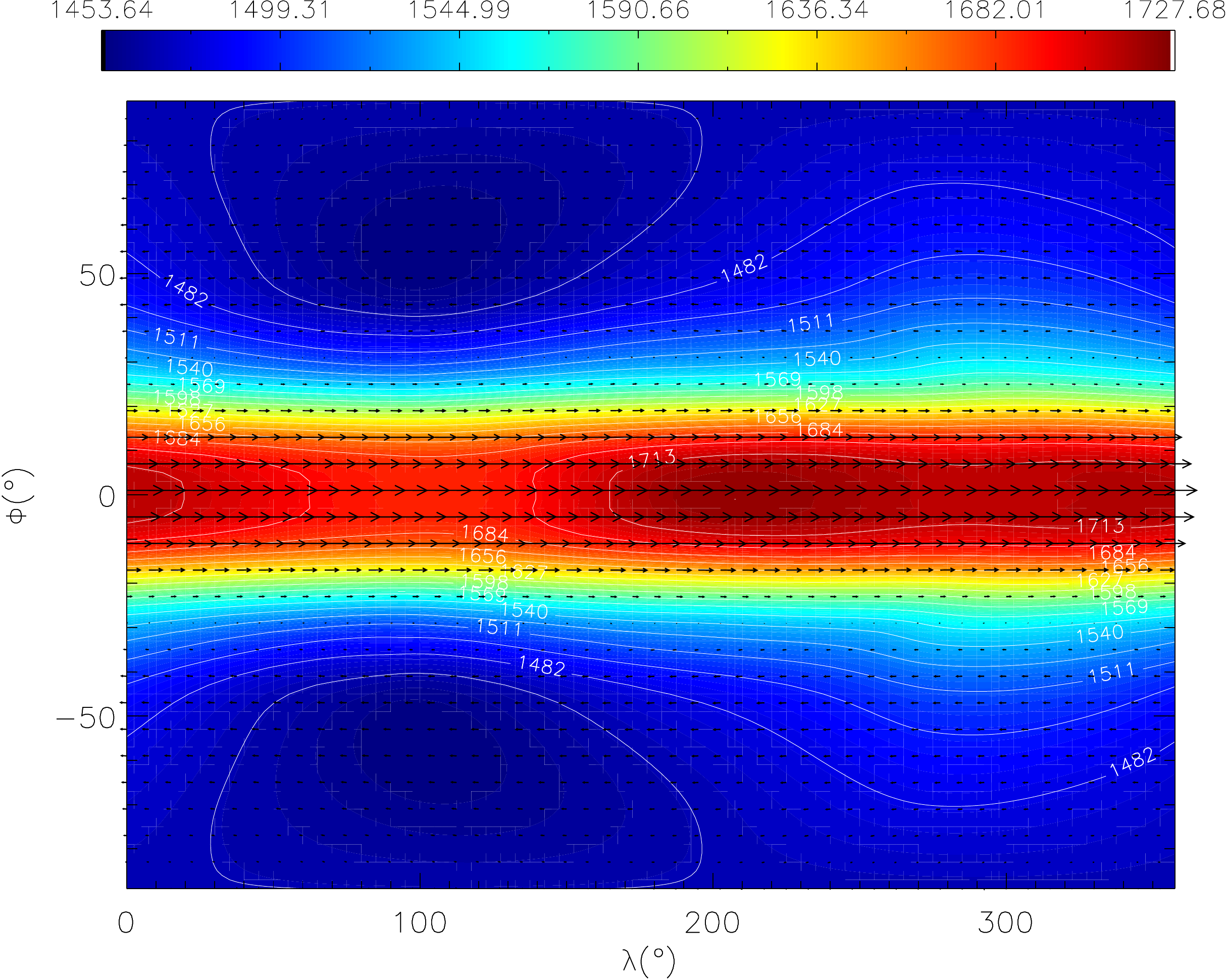}
\caption{The horizontal wind velocity as arrows and temperature [\si{\kelvin}] as colours and contours from our models of HD~209458b after $\SI{1600}{\day}$. Left column: Results from the model with Newtonian forcing discussed in \cref{sec:forcing} at \SIlist{100;3e3;3e4;1e5}{\pascal}~\citep[from top to bottom, from][]{Mayne2014a}. Right column: Results from the coupled model discussed in \cref{sec:coupled_um_ntv} at \SIlist{3;3e3;3e4;1e5}{\pascal} (from top to bottom).}
\label{fig:temp_wind}
\end{figure*}

We show in \cref{fig:temp_wind} (left column) the temperature and horizontal wind at various atmospheric depths as a function of longitude and latitude after \SI{1600}{\day}. At \SI{100}{\pascal} winds diverge from the hotspot located at substellar point ($\ang{180}$ longitude, $\ang{0}$ latitude). It is worth noting that, due to the very small radiative timescale at $\SI{100}{\pascal}$, the temperature is almost identical to the equilibrium temperature, which causes a large temperature contrast ($> \SI{900}{\kelvin}$) between the day and night side. For increasing pressures, the dynamical regime is dominated by a super-rotating equatorial jet spanning all longitudes. Dynamical processes redistributing the heat away from the substellar point become more dominant, which causes the temperature difference between the day and night side to decrease.

\begin{figure*}
\centering
\includegraphics[width=\figurewidth]{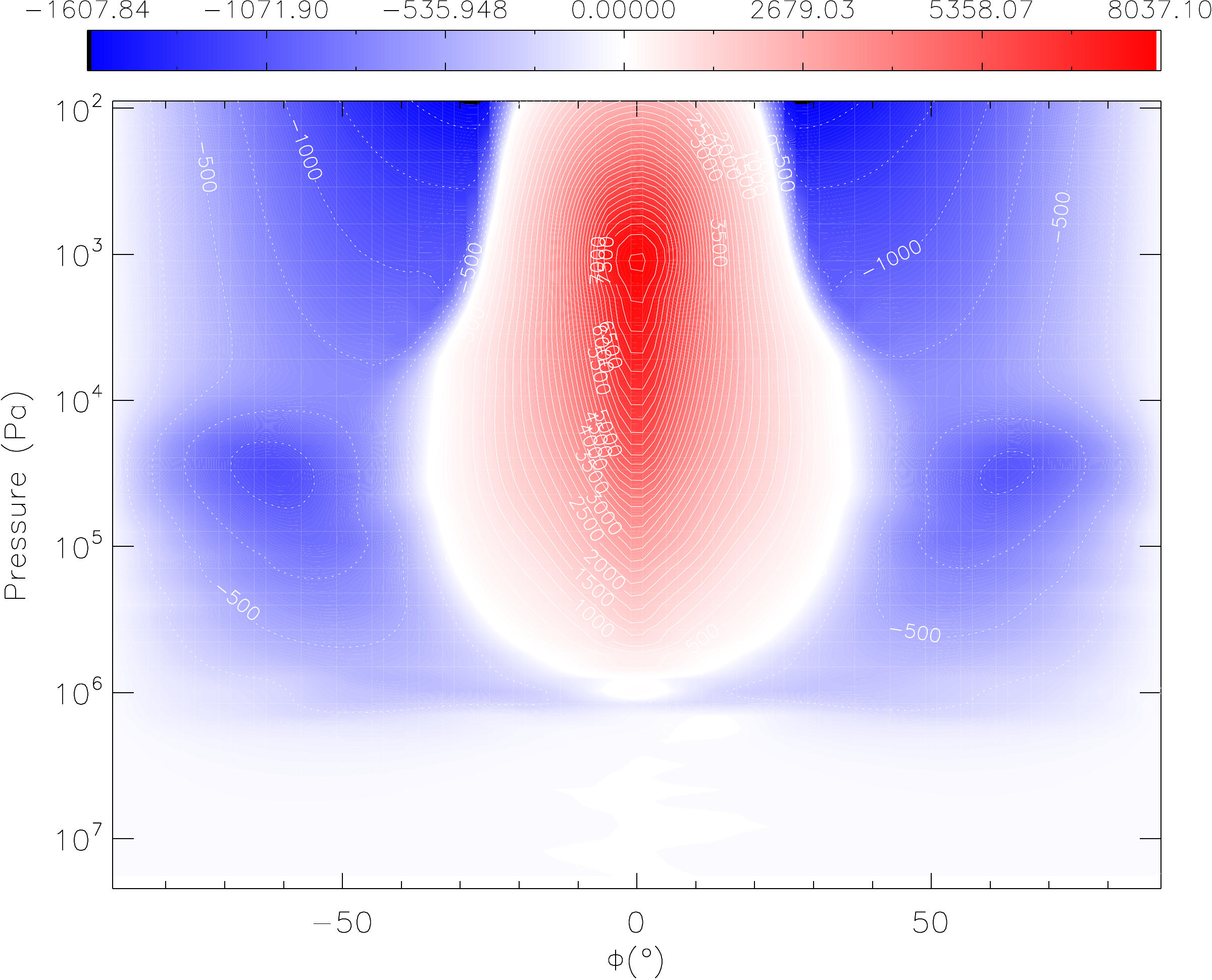}
\includegraphics[width=\figurewidth]{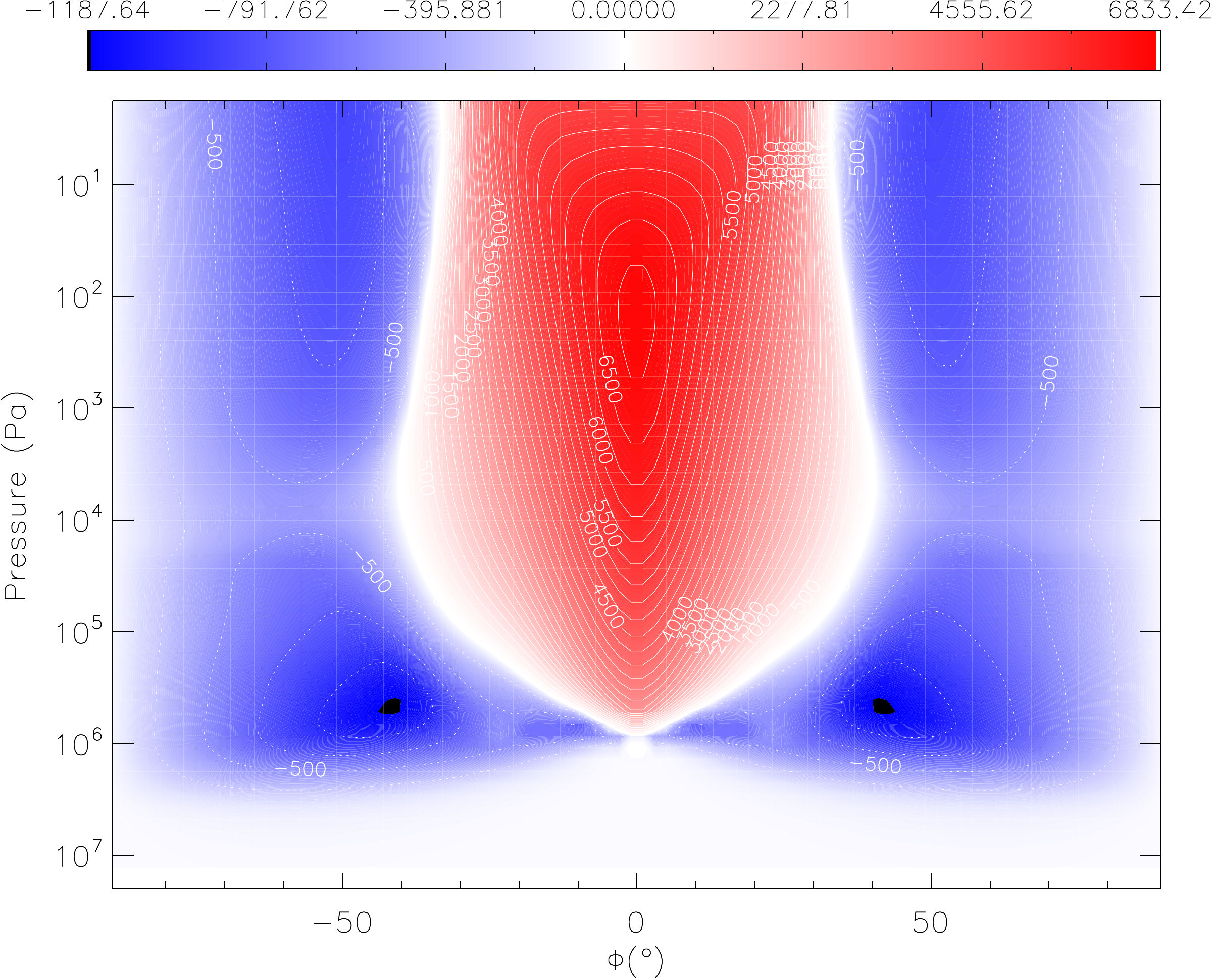}
\caption{The zonal mean of the zonal wind velocity [\si{\metre \per \second}] after $\SI{1600}{\day}$ for the models of HD~209458b. Left: The model with Newtonian forcing from \citet{Mayne2014a} discussed in \cref{sec:forcing}. Right: The coupled model discussed in \cref{sec:coupled_um_ntv}. Red indicates a prograde wind, blue indicates a retrograde wind.}
\label{fig:zonal_wind}
\end{figure*}

We show in the left-hand panel of \cref{fig:zonal_wind} the zonal mean of the zonal wind as a function of pressure and latitude. The zonal jet in the eastward direction mentioned above is clearly seen, and it reaches its maximum strength at about $\SI{e3}{\pascal}$ with a velocity of about $\SI{7}{\kilo \metre \per \second}$. At higher latitudes the mean flow is in the opposite (westward) direction, and much weaker in amplitude, with a maximum of about $\SI{1.2}{\kilo \metre \per \second}$.

\begin{figure}
\centering
\includegraphics[width=\figurewidth]{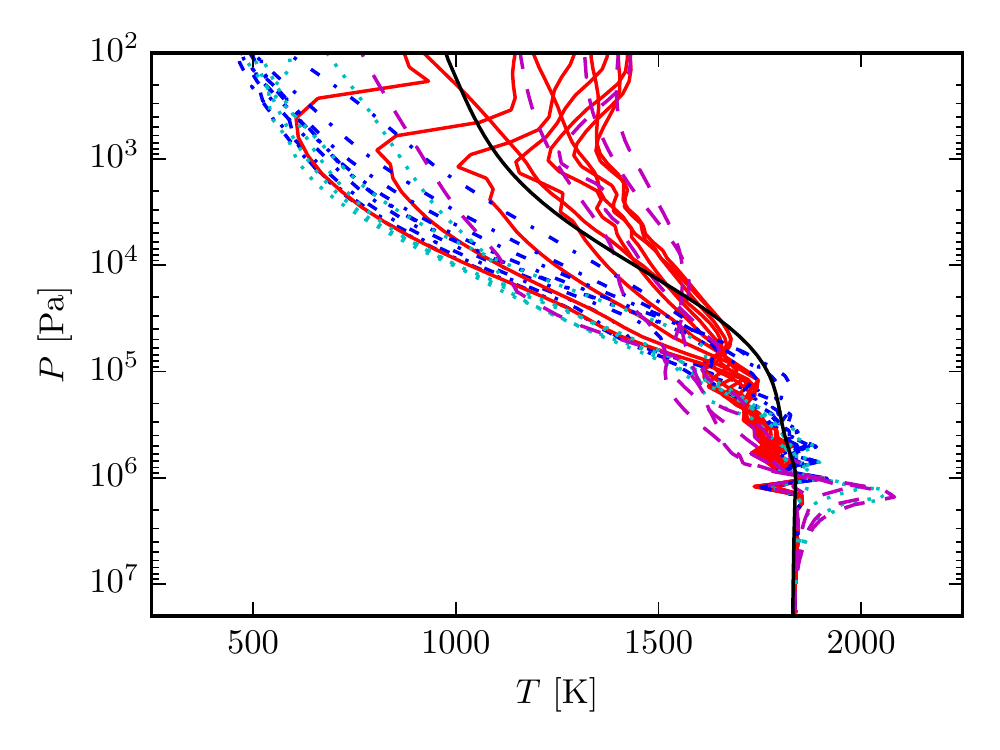}
\caption{$P$--$T$ profiles around the globe after $\SI{1600}{\day}$ for the model of HD~209458b with Newtonian forcing from \citet{Mayne2014a}. Red solid lines and blue dashed-dotted lines are day and night side profiles, respectively, at $\ang{0}$ latitude. Magenta dashed lines and cyan dotted lines are profiles between $\ang{0}$ and $\ang{90}$ latitude for longitudes $\ang{180}$ and $\ang{0}$, respectively.}
\label{fig:pt_forced}
\end{figure}

In \cref{fig:pt_forced} we plot $P$--$T$ profiles for several different latitudes and longitudes. The temperature varies significantly across the globe, with night side temperatures down to $\sim \SI{600}{\kelvin}$ and day side temperatures up to $\sim \SI{1500}{\kelvin}$ at $\SI{e3}{\pascal}$. A dynamically induced temperature inversion is even seen on the day side of the planet, which is caused by strong heating at the top of the atmosphere due to the short radiative timescale and the equatorial jet bringing cold material from the night side to the day side cooling the atmosphere down at larger pressures.

\subsection{Discussion} \label{sec:forcing_discussion}

\begin{figure}[t]
\centering
\includegraphics[width=\figurewidth]{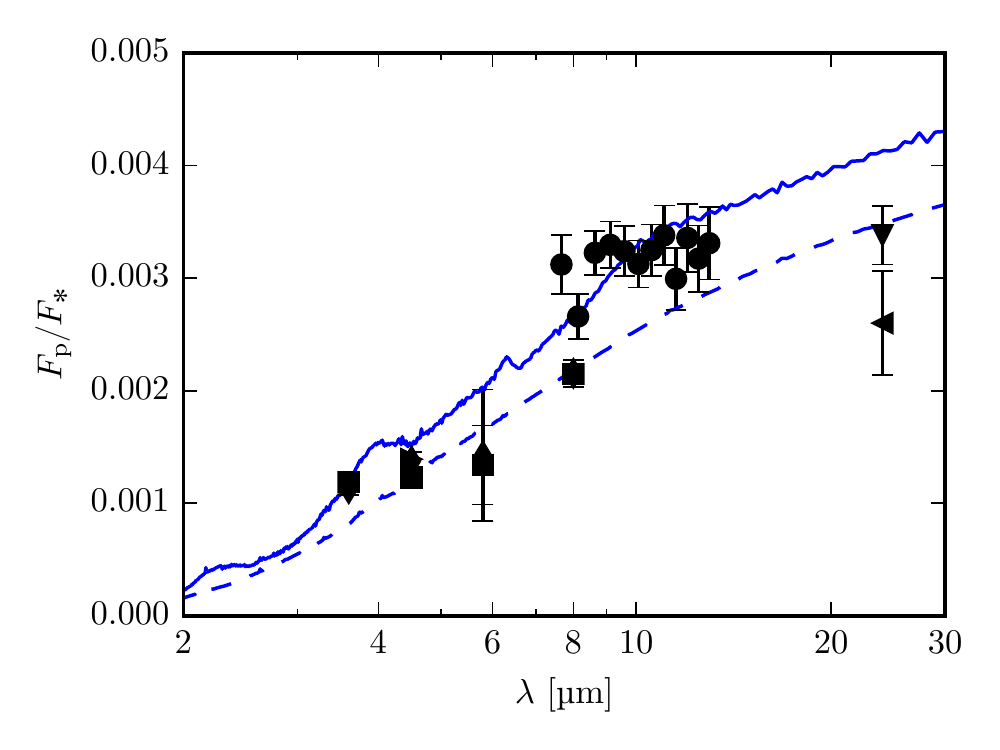}
\caption{Observed (points) and synthetic (lines) emission spectra for HD~209458b models with Newtonian cooling. The solid line has been calculated from the model presented in \citet{Mayne2014a} and \cref{sec:forcing} using \texttt{ATMO}, the dashed line is the synthetic emission spectrum from \citet{Fortney2006b}, which is based on the models of \citet{Cooper2005,Cooper2006} using the same Newtonian forcing scheme. The black points are observations from \citet{Swain2008} (\FilledCircle), \citet{Crossfield2012} (\FilledTriangleDown), \citet{Deming2005} (\FilledTriangleLeft), \citet{Zellem2014} (\FilledTriangleRight), \citet{Diamond-Lowe2014} (\FilledSquare) and \citet{Evans2015} (\FilledDiamondshape).}
\label{fig:emission_spec_tf}
\end{figure}

\begin{figure}
\centering
\includegraphics[width=\figurewidth]{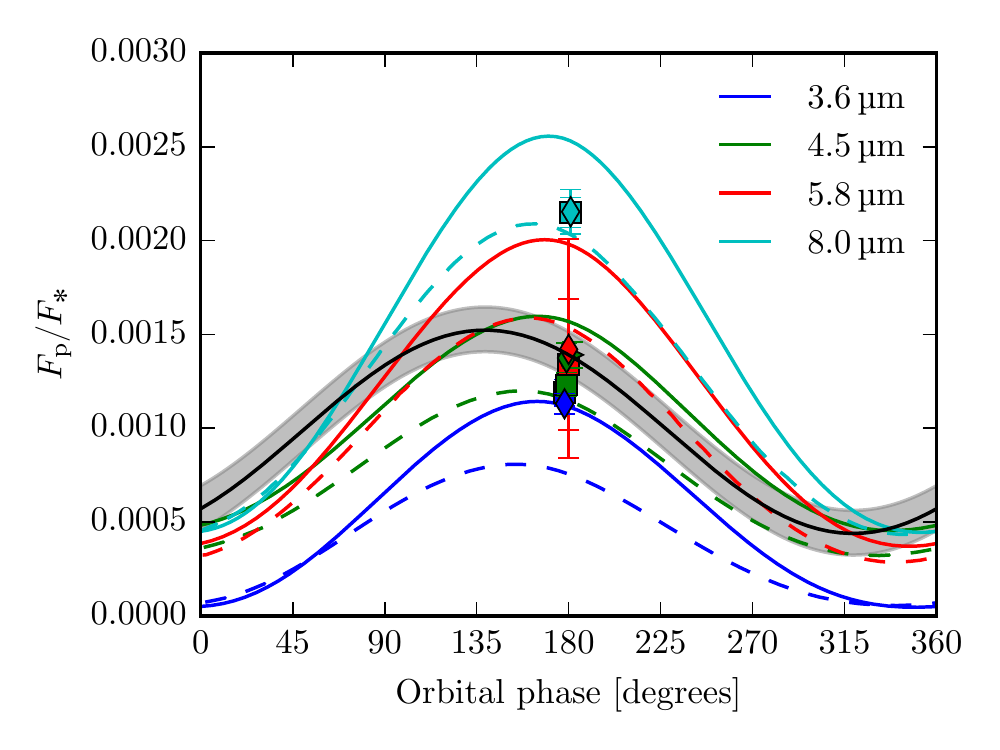}
\caption{Synthetic Spitzer IRAC phase curves from the Newtonian forcing model. The solid lines are calculated from the model presented in \citet{Mayne2014a} and \cref{sec:forcing} using \texttt{ATMO}, the dashed lines are the synthetic phase curves from \citet{Fortney2006b}, which are based on the models of \citet{Cooper2005,Cooper2006} using the same Newtonian forcing scheme. The models have been integrated over the IRAC bands using the filter functions. The data points are from \citet{Zellem2014} (\FilledTriangleRight), \citet{Diamond-Lowe2014} (\FilledSquare) and \citet{Evans2015} (\FilledDiamondshape). The best fit to the observed \SI{4.5}{\micro \metre} phase curve from \citet{Zellem2014} is shown as a solid black line, the grey shaded area is the $1\sigma$ uncertainty for the offset of the observed planet to star flux ratio.}
\label{fig:phase_curves_spitzer_tf}
\end{figure}

These results are similar to those obtained with other hot Jupiter GCMs using Newtonian forcing schemes~\citep[see e.g.][]{Cooper2005,Showman2008,Heng2011,Rauscher2010}, as discussed in \citet{Mayne2014a}. Using \texttt{ATMO} we have calculated synthetic day-side emission spectra and phase curves, which are shown in \cref{fig:emission_spec_tf,fig:phase_curves_spitzer_tf}, respectively, together with observational data points from the literature.

The day-side emission spectrum agrees reasonably well with observations. This is rather surprising as the forcing profiles were estimated from the globally averaged $P$--$T$ profile of \citet{Iro2005}~\citep{Cooper2005,Rauscher2010,Heng2011}, and are therefore not expected to be particularly accurate. The amplitudes of the \SI{4.5}{\micro \metre} phase curve show reasonably good agreement with the observed phase curve~\citep{Zellem2014}, but the significant phase offset in the observed phase curve is lacking. The latter could be due to an underestimate of the radiative timescale, which would lead to a smaller offset of the hottest point in the atmosphere from the substellar point.

\citet{Fortney2006b} presented synthetic day side emission spectra and phase curves using results from the GCM presented in \citet{Cooper2005,Cooper2006}, which uses the same forcing scheme as the one adopted here~\citep{Mayne2014a}. We have plotted the synthetic observations from \citet{Fortney2006b} in \cref{fig:emission_spec_tf,fig:phase_curves_spitzer_tf} as dashed lines to ease comparison with our synthetic observations. Our day-side emission is somewhat larger than that obtained by \citet{Fortney2006b}, while differences in night side fluxes are smaller. There is also a noticeable difference in the phase offsets of the peak flux from $\ang{180}$ between the models, we obtain a significantly smaller phase offset than \citet{Fortney2006b}.

It is difficult to pinpoint the exact causes of these differences, but there are several factors that might contribute: Slight discrepancies in the temperature may be caused by numerical details of the GCMs. The model of \citet{Cooper2005,Cooper2006} solve the primitive equations on a pressure based grid using a gravity constant with height, while we solve the full 3D Navier--Stokes equations with a height-varying gravity on a height-based grid. In addition the numerical schemes, grids and resolutions are different. In our model, however, solving the primitive equations and assuming a gravity constant with height only has a minor effect on the emission compared to solving the full 3D Navier--Stokes equations with a height-varying gravity.

The main differences in emission may therefore be caused by differences in the tools used for post-processing such as different line list and line width sources, and slightly different elemental abundances, resulting in the calculation of somewhat different opacities. The fact that GCMs with the same simplified forcing scheme give such different results emphasise the need to include the post-processing tools in intercomparison studies.

Before discussing results from the coupled model we briefly reiterate the disadvantages of using Newtonian forcing schemes, as in e.g. \citet{Cooper2005}, \citet{Heng2011}, \citet{Mayne2014a}, \citet{Menou2009}, \citet{Rauscher2010}, \citet{Showman2002} and \citet{Showman2008}, to treat the radiation:
\begin{enumerate}
\item
Equilibrium $P$--$T$ profiles needed by the forcing schemes, which vary as a function of latitude and longitude, are difficult to obtain from 1D models.
\item
The temperature relaxation is linear while in reality it is non-linear for large deviations from the equilibrium profiles.
\item
Atmospheric interactions due to exchange of radiative energy such as emission and absorption of thermal radiation are ignored.
\item
The model flexibility is poor since for each new planet modelled the forcing prescription must be changed.
\end{enumerate}
Global circulation models with Newtonian forcing schemes can still be useful for exploring underlying dynamical processes, see e.g. \citet{Showman2011b} and \citet{Komacek2016}. Including a proper treatment of radiative heating and cooling are essential, however, in order to improve model flexibility and, as we demonstrate in the next section (and discuss in \cref{sec:introduction}), improve agreement with observations.

\section{Full coupled model results} \label{sec:coupled_um_ntv}

Here we present results from simulations using the UM incorporating the full radiation scheme discussed in \citet{Amundsen2014}, including the modifications described in \cref{sec:radiation_scheme}. In the literature it is usually not explicitly stated how long simulations have been run for~\citep{Showman2009,Kataria2013,Kataria2014,Kataria2015}. We have run simulations for $> \SI{1000}{\day}$. For the observable part of the atmosphere, that is, pressures $\lesssim \SI{e5}{\pascal}$, the atmosphere has approximately reached a steady-state, while at larger pressures the atmosphere is still evolving and much longer integration timescales would be needed to study the evolution and its ramifications.

\subsection{Results} \label{sec:coupled_um_ntv_results}

Horizontal wind velocities and temperatures are plotted in \cref{fig:temp_wind} (right column) after $\SI{1600}{\day}$ at several different pressures, and can be compared to the left column obtained with the model with Newtonian forcing. General features are similar to those found in the model with Newtonian forcing. At low pressures ($\lesssim \SI{100}{\pascal}$) the flow is again diverging from the substellar point, with a hotspot shifted eastward from the substellar point.

In the right panel of \cref{fig:zonal_wind} we show the zonal mean zonal wind velocity after $\SI{1600}{\day}$ as a function of pressure and latitude. As in the model with Newtonian forcing the eastward equatorial jet is prominent, with a mean westward flow at higher latitudes.

\begin{figure}
\centering
\includegraphics[width=\figurewidth]{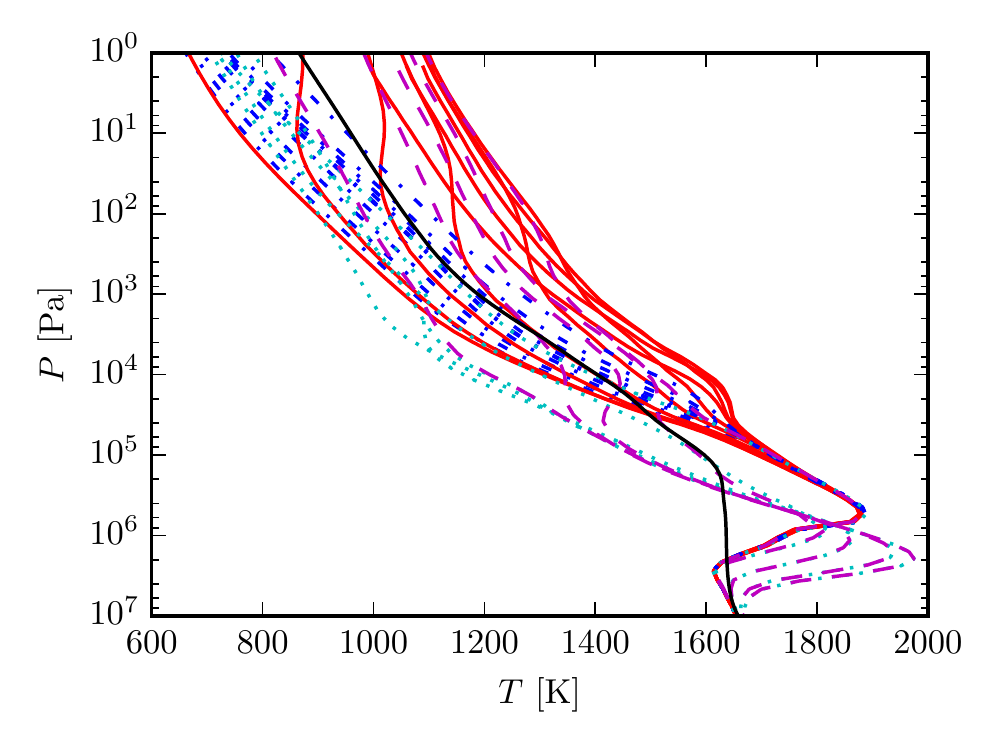}
\caption{$P$--$T$ profiles around the globe after $\SI{1600}{\day}$ for the coupled model of HD~209458b discussed in \cref{sec:coupled_um_ntv}. Red solid lines and blue dashed-dotted lines are day and night side profiles, respectively, at $\ang{0}$ latitude. Magenta dashed lines and cyan dotted lines are profiles between $\ang{0}$ and $\ang{90}$ latitude for longitudes $\ang{180}$ and $\ang{0}$, respectively. The black line is the initial $P$--$T$ profile adopted.}
\label{fig:pt_coupled_ntv}
\end{figure}

In \cref{fig:pt_coupled_ntv} we show the variation in $P$--$T$ profiles across the globe. A large variation is evident, and some night side profiles have higher temperatures than some day side profiles. This is due to the strong eastward advection causing the terminator at $\ang{270}$ longitude to be much warmer than that at $\ang{90}$ longitude. For pressures $\gg \SI{e5}{\pascal}$ profiles at $\ang{0}$ latitude are dominated by the equatorial jet, causing very small temperature variations as a function of longitude. At other latitudes, however, temperature variations are larger. The deep atmosphere between \num{e5} and \SI{e7}{\pascal} is generally much hotter than the initial $P$--$T$ profile, with temperatures approaching \SI{2000}{\kelvin}. This region is, due to the long dynamical timescale, evolving very slowly, and has not yet converged to a steady-state~\citep{Mayne2014a}. It is clear, however, that temperatures are slowly increasing across the globe at these deep levels.

\begin{figure*}
\centering
\includegraphics[width=\figurewidth]{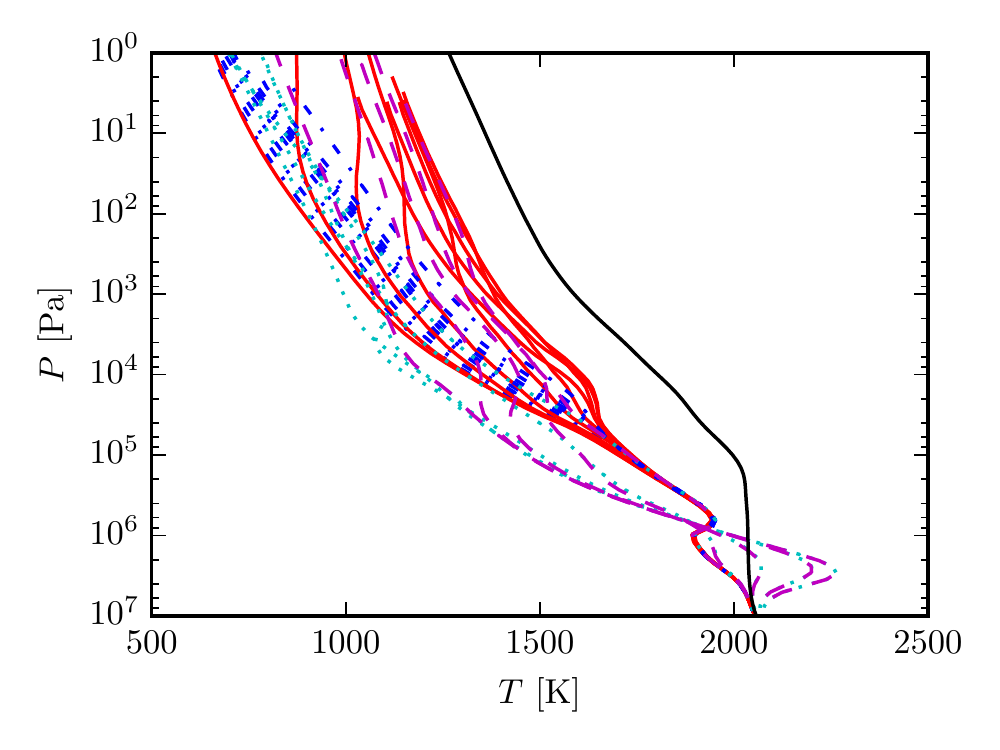}
\includegraphics[width=\figurewidth]{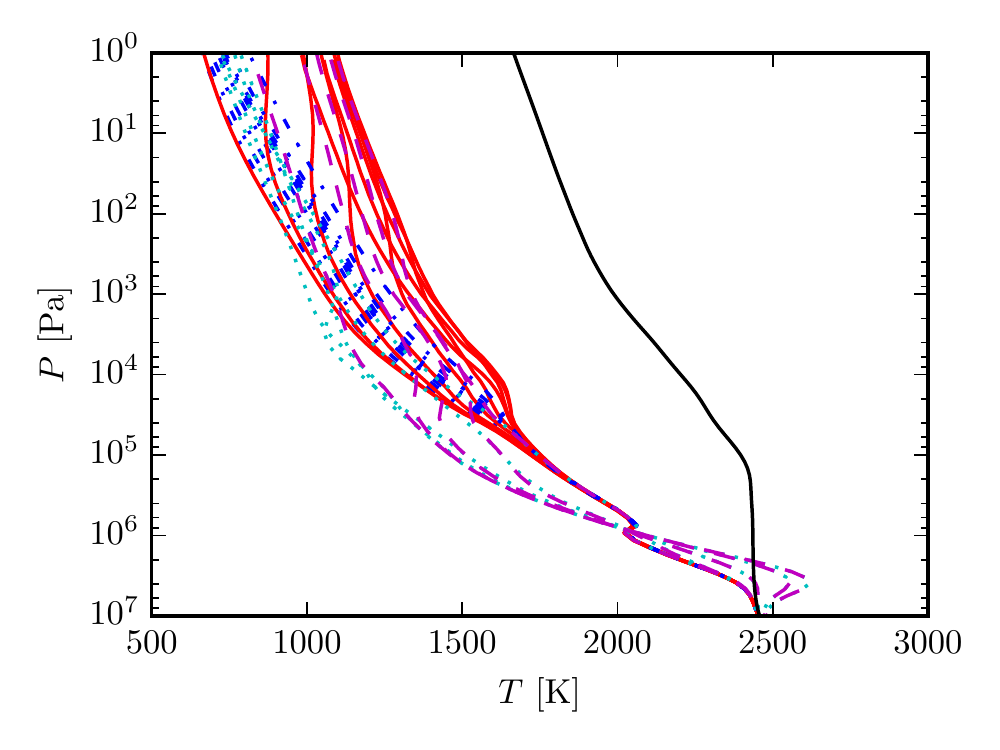}
\caption{$P$--$T$ profiles around the globe after $\SI{1600}{\day}$ for the coupled model of HD~209458b initialised with a $P$--$T$ profile that is \SI{400}{\kelvin} (left) and \SI{800}{\kelvin} (right) hotter than the global 1D mean. Lines are as in \cref{fig:pt_coupled_ntv}.}
\label{fig:pt_coupled_hot}
\end{figure*}

\begin{figure}
\centering
\includegraphics[width=\figurewidth]{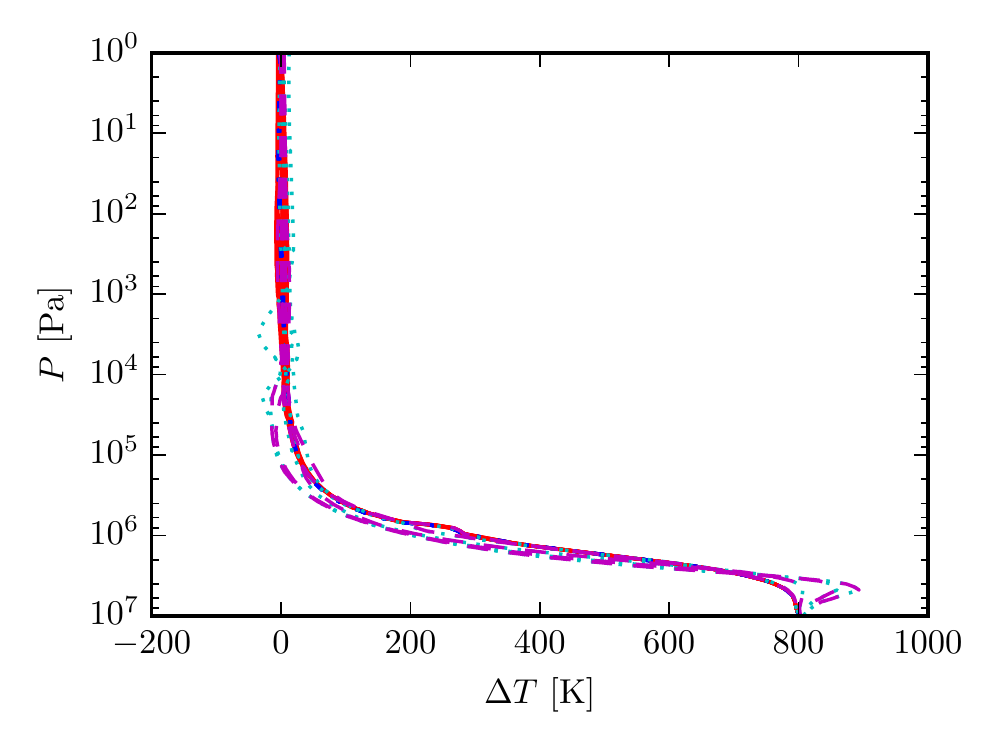}
\caption{Temperature difference $\Delta T$ between the $P$--$T$ profiles in the right-hand panel of \cref{fig:pt_coupled_hot}, which are from the model with a \SI{800}{\kelvin} hotter initial condition compared to the global 1D mean, and the $P$--$T$ profiles in \cref{fig:pt_coupled_ntv}, which are from the model initialised with the global 1D mean $P$--$T$ profile. Temperature differences are small for $P \lesssim \SI{e5}{\bar}$, while the models have clearly not reached a steady-state for $P \gtrsim \SI{e5}{\bar}$. Results are similar for the case with a \SI{400}{\kelvin} hotter initial condition. Lines are as in \cref{fig:pt_coupled_ntv}.}
\label{fig:pt_coupled_hot_diff}
\end{figure}

We also run simulations using a hotter initial temperature pressure profile, increasing from the standard global average 1D profile, uniformly, by \SI{400}{\kelvin} and \SI{800}{\kelvin}. The results are shown in \cref{fig:pt_coupled_hot,fig:pt_coupled_hot_diff}, where it becomes clear that the initial $P$--$T$ has not converged to a steady-state for $P \gtrsim \SI{e5}{\pascal}$. In fact, the atmosphere may be converging towards temperatures significantly hotter than estimated from a 1D global average at $P \sim \SI{e6}{\pascal}$. Unfortunately, due to computational limitations, we are unable to run our model for significantly longer timescales. However, these initial results suggest further work is required with regard to the sensitivity of hot Jupiter GCM results to such changes in the initial profile.

\subsection{Discussion} \label{sec:coupled_um_ntv_discussion}

Our results are in qualitative agreement with those of \citet{Showman2009}. The model exhibits a strong eastward equatorial jet, and hotspot shifted eastward of the substellar point. Unfortunately SPARC/MITgcm results for HD~209458b without TiO and VO have not been published in detail, which prohibits a more detailed comparison of temperature and wind fields. One characteristic of the hot Jupiter simulations presented in \citet{Showman2009} is what the authors term a ``vertical coherency'' of temperatures. This is particularly apparent in the solar metallicity HD~187733b setup, which excludes TiO and VO opacities, and is the most well matched to our HD~209458b simulations in terms of opacities. Vertical coherency describes the fact that the horizontal position of the hottest and coldest parts of the atmosphere vary only modestly between $\SI{e2}{\pascal}$ and $\SI{e5}{\pascal}$. This was not seen, nor expected, in previous simulations adopting Newtonian forcing, as the radiative timescale varies by about two orders of magnitude over these depths. Therefore, one might expect the balance between the radiative forcing and advection to change with depth and lead to a significant change in the horizontal temperature distribution. \citet{Showman2009} proposed that the observed vertical coherency was caused by the vertical interaction of thermal radiation reducing vertical temperature gradients. This effect is self-consistently included in the models of \citet{Showman2009}, but not included in those adopting Newtonian forcing.

We do not see vertical coherency in our simulations, despite self-consistently treating the thermal radiation: the position of the hottest and coldest points vary significantly with pressure. This is particularly noticeable at $\SI{e5}{\pascal} = \SI{1}{\bar}$ where both winds and temperatures are dominated by the eastward equatorial jet, and a weak retrograde flow at higher latitudes. Our longitudinal temperature variations are very small ($< \SI{100}{\kelvin}$), in contrast to the models in \citet{Showman2009} where temperatures vary by up to \SI{500}{\kelvin} at high latitudes. The reason for this discrepancy is unclear, but we have run our model significantly longer, giving the system time to equilibrate at higher pressures, and we do not assume the atmosphere to be shallow. This may help explain the weaker vertical coherency in our model, but more in-depth comparisons are needed to understand these differences in more detail.

The fact that the deep layers are heating up compared to the initial $P$--$T$ profile is intriguing. Even though the temperatures have not converged to a steady-state, it suggests that the deep atmosphere, in equilibrium, should be hotter than predicted by simple 1D models. Further analysis is required to understand this feature.

\begin{figure}
\centering
\includegraphics[width=\figurewidth]{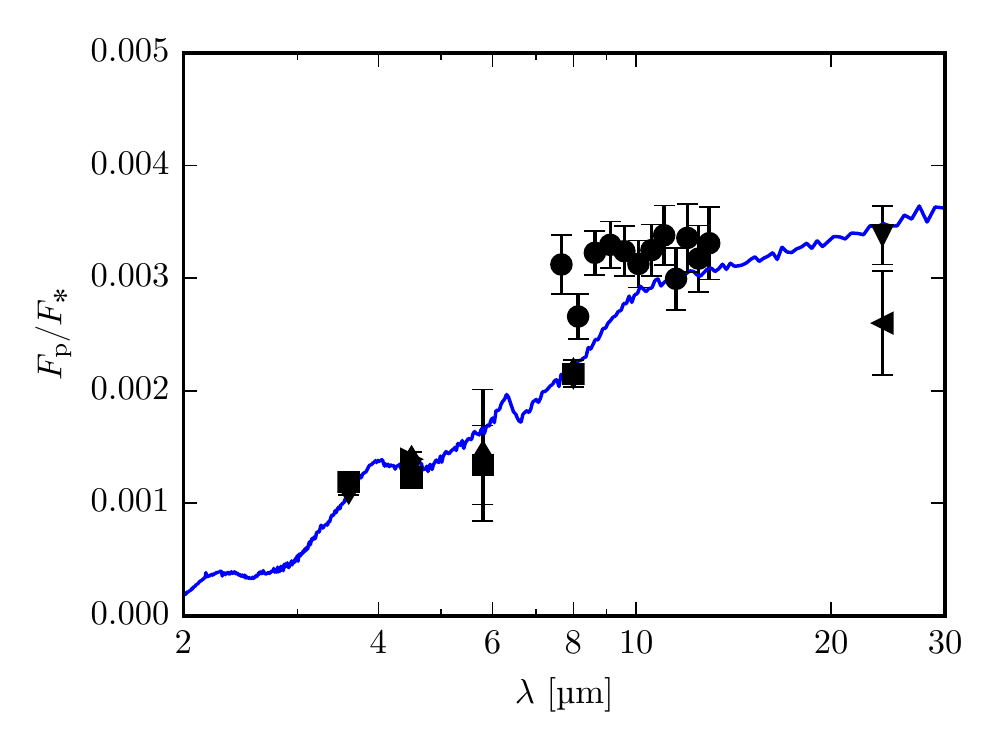}
\caption{Same as \cref{fig:emission_spec_tf}, but the synthetic emission spectrum has been calculated using results from our coupled model (\cref{sec:coupled_um_ntv}) with \texttt{ATMO}.}
\label{fig:emission_spec_ntv}
\end{figure}

\begin{figure}
\centering
\includegraphics[width=\figurewidth]{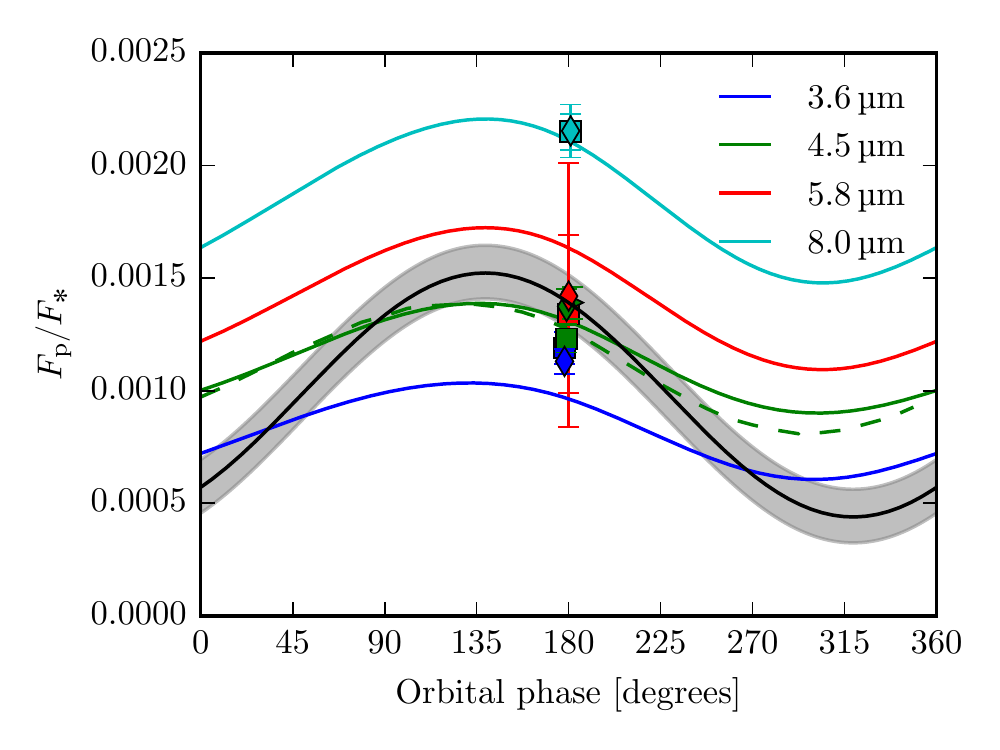}
\caption{Same as \cref{fig:phase_curves_spitzer_tf}, but the synthetic emission spectrum has been calculated using results from our coupled model (\cref{sec:coupled_um_ntv}) with \texttt{ATMO}. The dashed line is the model \SI{4.5}{\micro \metre} phase curve from \citet{Zellem2014} obtained using the GCM of \citet{Showman2009} without TiO and VO.}
\label{fig:phase_curves_spitzer_ntv}
\end{figure}

We show in \cref{fig:emission_spec_ntv,fig:phase_curves_spitzer_ntv} the synthetic day side emission spectrum and phase curves calculated using \texttt{ATMO}. The day side emission spectrum provides a reasonably good fit to the observations, and so does the offset of the peak emission from $\ang{180}$ in the $\SI{4.5}{\micro \metre}$ phase curve, while the offset of the minimum emission is larger than observed.

As mentioned above, \citet{Showman2009} do not provide results for their model of HD~209458b without TiO and VO, which prevents direct comparison. The \SI{4.5}{\micro \metre} phase curve from their model without TiO and VO is, however, presented in \citet{Zellem2014}, and we have plotted it in \cref{fig:phase_curves_spitzer_ntv} as a dashed line. The model \SI{4.5}{\micro \metre} phase curves agree reasonably well, with minor differences in the night side emission. Interestingly, both models significantly overestimate the night side flux, indicating that this is a common feature of current GCMs. \citet{Zellem2014} suggest that this may be due to non-equilibrium carbon chemistry, specifically vertical quenching of CH$_4$, leading to larger CH$_4$ abundances and consequently more efficient night side cooling. Another potential explanation could be horizontal quenching of CO increasing the abundance of CO on the night side relative to that predicted by equilibrium. As CO has a strong absorption feature at $\SI{4.5}{\micro \metre}$ this could potentially decrease the emission at this wavelength.

We have previously found that the effect of non-equilibrium chemistry has likely been overestimated in previous studies due to the inconsistent treatment of non-equilibrium chemistry and its feedback on the $P$--$T$ profile~\cite{Drummond2016}. In addition, most previous studies are limited to considering vertical non-equilibrium effects due to the 1D models used. To investigate the effect of horizontal mixing in these atmospheres, which could dominate over vertical mixing due to the large horizontal wind speeds, we are currently coupling a non-equilibrium chemistry network to the GCM, which may provide a solution to our current inability to reproduce the observed night side emission.

The lack of a temperature inversion potentially caused by TiO and VO in the atmosphere of HD~209458b has been suggested to be due to a cold-trap~\citep{Hubeny2003,Spiegel2009,Parmentier2013,Parmentier2016}. TiO and VO could potentially be advected from the day side to the night side, condense, rain out and be trapped on night side at high pressures. We note that, as these calculations were based on GCM results where the model, like ours, has not equilibrated for pressures $P \gtrsim \SI{e5}{\pascal}$. Temperatures at these pressures may be higher than estimated by the 1D global average used to initialise the models, see \cref{fig:pt_coupled_ntv,fig:pt_coupled_hot}, and we therefore emphasise the need to study the long-term evolution of the deep atmosphere of hot Jupiters, both to study potential cold-traps and radius inflation.

We note that all temperatures in our model initialised with the 1D globally averaged $P$--$T$ profile are below the condensation temperature of TiO/VO. Consequently, we would not expect gaseous TiO and VO to form even if we included their opacity. They may form, however, shortly after the simulations start as the models are initialised with zero advection, causing the day side to initially heat up to temperatures potentially above the condensation temperature, and then cool down as the advection becomes more efficient. Studying the formation of TiO and VO with varying initial conditions in these models would therefore be beneficial in order to understand the robustness of the presence of these gases in hot Jupiter atmospheres.

\section{Conclusions} \label{sec:conclusions}

We have presented results from the first application of the UM, including a sophisticated and accurate radiation scheme, to hot Jupiters. We have performed comparisons to both the model of \citet{Mayne2014a} which employs Newtonian forcing, to the SPARC/MITgcm of \citet{Showman2009} and to existing observations. Our main conclusions are
\begin{itemize}
\item
Our models with Newtonian forcing and sophisticated radiative transfer have qualitatively similar wind and temperature patterns: a hotspot shifted eastward of the substellar point, a broad eastward equatorial jet and a westward mean flow at higher latitudes. This is in good qualitative agreement with results obtained with the SPARC/MITgcm \citep{Showman2009}.
\item
The day side emission and phase offset of the coupled model fit the observed \SI{4.5}{\micro \metre} phase curve reasonably well, while the night side emission is about a factor of two too large. This is similar to the results of \citet{Showman2009}, and potential explanations are effects of non-equilibrium chemistry and super-solar metallicities. We are in the process of investigating this by coupling a non-equilibrium chemistry scheme to our GCM.
\item
We do not see the vertical coherence of temperatures, that is, the small variation in the location of the hottest and coldest points in the atmosphere with pressure, seen in the SPARC/MITgcm. The cause of this difference is unclear, but the vertical coherence is not expected from simple time scale arguments.
\item
The deep atmosphere has not converged to a steady-state even though we have run the model for \SI{1600}{\day}, and we see evidence for significant deviations from the globally averaged 1D radiative-convective equilibrium profile used to initialise the model. This highlights the need to use very long runs when evaluating the feasibility of cold-traps and studying mechanisms for radius inflation.
\item
The observed differences between the UM and the SPARC/MITgcm highlight the importance of model intercomparisons, which are needed to separate physically robust results from model degeneracies. There is a need to compare the dynamical cores in more detail, but also both the radiation schemes and post-processing tools in isolation. In addition it will be important study the sensitivity of the model to the initialisation, particularly when including strong absorbers of stellar radiation with high condensation temperatures such as TiO and VO.
\end{itemize}

Despite the difficulties discussed here, GCMs have seen rapid improvement in their application to hot Jupiters. Combined with ever improving observations, as well as extensive intercomparison exercises, they will enable us to significantly improve our understanding of these exotic planets.

\begin{acknowledgements}
We would like to thank Jonathan Tennyson and Travis Barman for insightful discussions. This work is partly supported by the European Research Council under the European Community's Seventh Framework Programme (FP7/2007-2013 Grant Agreement No. 247060-PEPS and grant No. 320478-TOFU). DSA acknowledges support from the NASA Astrobiology Program through the Nexus for Exoplanet System Science. NM acknowledges funding from the Leverhulme Trust via a Research Project Grant. JM and CS acknowledge the support of a Met Office Academic Partnership secondment. DH acknowledges funding from the DFG through the Collaborative Research Centre SFB 881 ``The Milky Way System''. The calculations for this paper were performed on the University of Exeter Supercomputer, a DiRAC Facility jointly funded by STFC, the Large Facilities Capital Fund of BIS, and the University of Exeter.
\end{acknowledgements}

\bibliographystyle{aa}
\bibliography{bibliography_hot_Jupiters}

\begin{thebibliography}{90}
\expandafter\ifx\csname natexlab\endcsname\relax\def\natexlab#1{#1}\fi

\bibitem[{{Ag{\'u}ndez} {et~al.}(2014){Ag{\'u}ndez}, {Parmentier}, {Venot},
  {Hersant}, \& {Selsis}}]{Agundez2014}
{Ag{\'u}ndez}, M., {Parmentier}, V., {Venot}, O., {Hersant}, F., \& {Selsis},
  F. 2014, \aap, 564, A73

\bibitem[{{Allard} {et~al.}(2003){Allard}, {Allard}, {Hauschildt}, {Kielkopf},
  \& {Machin}}]{Allard2003}
{Allard}, N.~F., {Allard}, F., {Hauschildt}, P.~H., {Kielkopf}, J.~F., \&
  {Machin}, L. 2003, \aap, 411, L473

\bibitem[{{Allard} {et~al.}(1999){Allard}, {Royer}, {Kielkopf}, \&
  {Feautrier}}]{Allard1999}
{Allard}, N.~F., {Royer}, A., {Kielkopf}, J.~F., \& {Feautrier}, N. 1999, \pra,
  60, 1021

\bibitem[{{Allard} {et~al.}(2007){Allard}, {Spiegelman}, \&
  {Kielkopf}}]{Allard2007}
{Allard}, N.~F., {Spiegelman}, F., \& {Kielkopf}, J.~F. 2007, \aap, 465, 1085

\bibitem[{{Amundsen}(2015)}]{Amundsen2015thesis}
{Amundsen}, D.~S. 2015, PhD thesis, {University of Exeter}

\bibitem[{{Amundsen} {et~al.}(2014){Amundsen}, {Baraffe}, {Tremblin},
  {Manners}, {Hayek}, {Mayne}, \& {Acreman}}]{Amundsen2014}
{Amundsen}, D.~S., {Baraffe}, I., {Tremblin}, P., {et~al.} 2014, \aap, 564, A59

\bibitem[{{Amundsen} {et~al.}(2016, submitted){Amundsen}, {Tremblin},
  {Manners}, {Baraffe}, \& {Mayne}}]{Amundsen2016b}
{Amundsen}, D.~S., {Tremblin}, P., {Manners}, J., {Baraffe}, I., \& {Mayne},
  N.~J. 2016, submitted, \aap

\bibitem[{{Burrows} \& {Sharp}(1999)}]{Burrows1999}
{Burrows}, A. \& {Sharp}, C.~M. 1999, \apj, 512, 843

\bibitem[{{Charbonneau} {et~al.}(2002){Charbonneau}, {Brown}, {Noyes}, \&
  {Gilliland}}]{Charbonneau2002}
{Charbonneau}, D., {Brown}, T.~M., {Noyes}, R.~W., \& {Gilliland}, R.~L. 2002,
  \apj, 568, 377

\bibitem[{{Collins} {et~al.}(2006){Collins}, {Ramaswamy}, {Schwarzkopf}, {Sun},
  {Portmann}, {Fu}, {Casanova}, {Dufresne}, {Fillmore}, {Forster}, {Galin},
  {Gohar}, {Ingram}, {Kratz}, {Lefebvre}, {Li}, {Marquet}, {Oinas}, {Tsushima},
  {Uchiyama}, \& {Zhong}}]{Collins2006}
{Collins}, W.~D., {Ramaswamy}, V., {Schwarzkopf}, M.~D., {et~al.} 2006, Journal
  of Geophysical Research (Atmospheres), 111, 14317

\bibitem[{{Cooper} \& {Showman}(2005)}]{Cooper2005}
{Cooper}, C.~S. \& {Showman}, A.~P. 2005, \apjl, 629, L45

\bibitem[{{Cooper} \& {Showman}(2006)}]{Cooper2006}
{Cooper}, C.~S. \& {Showman}, A.~P. 2006, \apj, 649, 1048

\bibitem[{{Crossfield} {et~al.}(2012){Crossfield}, {Knutson}, {Fortney},
  {Showman}, {Cowan}, \& {Deming}}]{Crossfield2012}
{Crossfield}, I.~J.~M., {Knutson}, H., {Fortney}, J., {et~al.} 2012, \apj, 752,
  81

\bibitem[{{Deming} {et~al.}(2005){Deming}, {Seager}, {Richardson}, \&
  {Harrington}}]{Deming2005}
{Deming}, D., {Seager}, S., {Richardson}, L.~J., \& {Harrington}, J. 2005,
  \nat, 434, 740

\bibitem[{{Deming} {et~al.}(2013){Deming}, {Wilkins}, {McCullough}, {Burrows},
  {Fortney}, {Agol}, {Dobbs-Dixon}, {Madhusudhan}, {Crouzet}, {Desert},
  {Gilliland}, {Haynes}, {Knutson}, {Line}, {Magic}, {Mandell}, {Ranjan},
  {Charbonneau}, {Clampin}, {Seager}, \& {Showman}}]{Deming2013}
{Deming}, D., {Wilkins}, A., {McCullough}, P., {et~al.} 2013, \apj, 774, 95

\bibitem[{{Diamond-Lowe} {et~al.}(2014){Diamond-Lowe}, {Stevenson}, {Bean},
  {Line}, \& {Fortney}}]{Diamond-Lowe2014}
{Diamond-Lowe}, H., {Stevenson}, K.~B., {Bean}, J.~L., {Line}, M.~R., \&
  {Fortney}, J.~J. 2014, \apj, 796, 66

\bibitem[{{Dobbs-Dixon} \& {Agol}(2013)}]{Dobbs-Dixon2013}
{Dobbs-Dixon}, I. \& {Agol}, E. 2013, \mnras, 435, 3159

\bibitem[{{Drummond} {et~al.}(2016, accepted){Drummond}, {Tremblin}, {Baraffe},
  {Amundsen}, {Mayne}, {Venot}, \& {Goyal}}]{Drummond2016}
{Drummond}, B., {Tremblin}, P., {Baraffe}, I., {et~al.} 2016, accepted, \aap

\bibitem[{{Edwards}(1996)}]{Edwards1996b}
{Edwards}, J.~M. 1996, Journal of Atmospheric Sciences, 53, 1921

\bibitem[{{Edwards} \& {Slingo}(1996)}]{Edwards1996a}
{Edwards}, J.~M. \& {Slingo}, A. 1996, Quarterly Journal of the Royal
  Meteorological Society, 122, 689

\bibitem[{{Ellingson} {et~al.}(1991){Ellingson}, {Ellis}, \&
  {Fels}}]{Ellingson1991}
{Ellingson}, R.~G., {Ellis}, J., \& {Fels}, S. 1991, \jgr, 96, 8929

\bibitem[{{Evans} {et~al.}(2015){Evans}, {Aigrain}, {Gibson}, {Barstow},
  {Amundsen}, {Tremblin}, \& {Mourier}}]{Evans2015}
{Evans}, T.~M., {Aigrain}, S., {Gibson}, N., {et~al.} 2015, \mnras, 451, 680

\bibitem[{{Evans} {et~al.}(2016){Evans}, {Sing}, {Wakeford}, {Nikolov},
  {Ballester}, {Drummond}, {Kataria}, {Gibson}, {Amundsen}, \&
  {Spake}}]{Evans2016}
{Evans}, T.~M., {Sing}, D.~K., {Wakeford}, H.~R., {et~al.} 2016, \apjl, 822, L4

\bibitem[{{Fortney} {et~al.}(2006){Fortney}, {Cooper}, {Showman}, {Marley}, \&
  {Freedman}}]{Fortney2006b}
{Fortney}, J.~J., {Cooper}, C.~S., {Showman}, A.~P., {Marley}, M.~S., \&
  {Freedman}, R.~S. 2006, \apj, 652, 746

\bibitem[{{Fortney} {et~al.}(2010){Fortney}, {Shabram}, {Showman}, {Lian},
  {Freedman}, {Marley}, \& {Lewis}}]{Fortney2010}
{Fortney}, J.~J., {Shabram}, M., {Showman}, A.~P., {et~al.} 2010, \apj, 709,
  1396

\bibitem[{{Heiter} {et~al.}(2008){Heiter}, {Barklem}, {Fossati}, {Kildiyarova},
  {Kochukhov}, {Kupka}, {Obbrugger}, {Piskunov}, {Plez}, {Ryabchikova},
  {Stempels}, {St{\"u}tz}, \& {Weiss}}]{Heiter2008}
{Heiter}, U., {Barklem}, P., {Fossati}, L., {et~al.} 2008, Journal of Physics
  Conference Series, 130, 012011

\bibitem[{{Heller}(1965)}]{Heller1965}
{Heller}, W. 1965, The Journal of Physical Chemistry, 69, 1123

\bibitem[{{Helling} {et~al.}(2016){Helling}, {Lee}, {Dobbs-Dixon}, {Mayne},
  {Amundsen}, {Khaimova}, {Unger}, {Manners}, {Acreman}, \&
  {Smith}}]{Helling2016}
{Helling}, C., {Lee}, G., {Dobbs-Dixon}, I., {et~al.} 2016, \mnras

\bibitem[{{Heng} {et~al.}(2011){Heng}, {Menou}, \& {Phillipps}}]{Heng2011}
{Heng}, K., {Menou}, K., \& {Phillipps}, P.~J. 2011, \mnras, 413, 2380

\bibitem[{{Hoeijmakers} {et~al.}(2014){Hoeijmakers}, {de Kok}, {Snellen},
  {Brogi}, {Birkby}, \& {Schwarz}}]{Hoeijmakers2014}
{Hoeijmakers}, H.~J., {de Kok}, R.~J., {Snellen}, I.~A.~G., {et~al.} 2014,
  ArXiv e-prints

\bibitem[{{Hubeny} {et~al.}(2003){Hubeny}, {Burrows}, \&
  {Sudarsky}}]{Hubeny2003}
{Hubeny}, I., {Burrows}, A., \& {Sudarsky}, D. 2003, \apj, 594, 1011

\bibitem[{{Iro} {et~al.}(2005){Iro}, {B{\'e}zard}, \& {Guillot}}]{Iro2005}
{Iro}, N., {B{\'e}zard}, B., \& {Guillot}, T. 2005, \aap, 436, 719

\bibitem[{{Kataria} {et~al.}(2014){Kataria}, {Showman}, {Fortney}, {Marley}, \&
  {Freedman}}]{Kataria2014}
{Kataria}, T., {Showman}, A.~P., {Fortney}, J.~J., {Marley}, M.~S., \&
  {Freedman}, R.~S. 2014, \apj, 785, 92

\bibitem[{{Kataria} {et~al.}(2015){Kataria}, {Showman}, {Fortney}, {Stevenson},
  {Line}, {Kreidberg}, {Bean}, \& {D{\'e}sert}}]{Kataria2015}
{Kataria}, T., {Showman}, A.~P., {Fortney}, J.~J., {et~al.} 2015, \apj, 801, 86

\bibitem[{{Kataria} {et~al.}(2013){Kataria}, {Showman}, {Lewis}, {Fortney},
  {Marley}, \& {Freedman}}]{Kataria2013}
{Kataria}, T., {Showman}, A.~P., {Lewis}, N.~K., {et~al.} 2013, \apj, 767, 76

\bibitem[{{Kataria} {et~al.}(2016){Kataria}, {Sing}, {Lewis}, {Visscher},
  {Showman}, {Fortney}, \& {Marley}}]{Kataria2016}
{Kataria}, T., {Sing}, D.~K., {Lewis}, N.~K., {et~al.} 2016, \apj, 821, 9

\bibitem[{{Knutson} {et~al.}(2007){Knutson}, {Charbonneau}, {Allen}, {Fortney},
  {Agol}, {Cowan}, {Showman}, {Cooper}, \& {Megeath}}]{Knutson2007a}
{Knutson}, H.~A., {Charbonneau}, D., {Allen}, L.~E., {et~al.} 2007, \nat, 447,
  183

\bibitem[{{Knutson} {et~al.}(2009){Knutson}, {Charbonneau}, {Cowan}, {Fortney},
  {Showman}, {Agol}, {Henry}, {Everett}, \& {Allen}}]{Knutson2009}
{Knutson}, H.~A., {Charbonneau}, D., {Cowan}, N.~B., {et~al.} 2009, \apj, 690,
  822

\bibitem[{{Knutson} {et~al.}(2012){Knutson}, {Lewis}, {Fortney}, {Burrows},
  {Showman}, {Cowan}, {Agol}, {Aigrain}, {Charbonneau}, {Deming}, {D{\'e}sert},
  {Henry}, {Langton}, \& {Laughlin}}]{Knutson2012}
{Knutson}, H.~A., {Lewis}, N., {Fortney}, J.~J., {et~al.} 2012, \apj, 754, 22

\bibitem[{{Komacek} \& {Showman}(2016)}]{Komacek2016}
{Komacek}, T.~D. \& {Showman}, A.~P. 2016, \apj, 821, 16

\bibitem[{{Lacis} \& {Oinas}(1991)}]{Lacis1991}
{Lacis}, A.~A. \& {Oinas}, V. 1991, \jgr, 96, 9027

\bibitem[{{Lee} {et~al.}(2016){Lee}, {Dobbs-Dixon}, {Helling}, {Bognar}, \&
  {Woitke}}]{Lee2016}
{Lee}, G., {Dobbs-Dixon}, I., {Helling}, C., {Bognar}, K., \& {Woitke}, P.
  2016, ArXiv e-prints

\bibitem[{{Leonard}(1974)}]{Leonard1974}
{Leonard}, P.~J. 1974, Atomic Data and Nuclear Data Tables, 14, 21

\bibitem[{{Lewis} {et~al.}(2010){Lewis}, {Showman}, {Fortney}, {Marley},
  {Freedman}, \& {Lodders}}]{Lewis2010}
{Lewis}, N.~K., {Showman}, A.~P., {Fortney}, J.~J., {et~al.} 2010, \apj, 720,
  344

\bibitem[{{Line} {et~al.}(2016){Line}, {Stevenson}, {Bean}, {Desert},
  {Fortney}, {Kreidberg}, {Madhusudhan}, {Showman}, \&
  {Diamond-Lowe}}]{Line2016}
{Line}, M.~R., {Stevenson}, K.~B., {Bean}, J., {et~al.} 2016, ArXiv e-prints

\bibitem[{{Louden} \& {Wheatley}(2015)}]{Louden2015}
{Louden}, T. \& {Wheatley}, P.~J. 2015, \apjl, 814, L24

\bibitem[{{Mansfield} \& {Peck}(1969)}]{Mansfield1969}
{Mansfield}, C.~R. \& {Peck}, E.~R. 1969, Journal of the Optical Society of
  America (1917-1983), 59, 199

\bibitem[{{Maxted} {et~al.}(2013){Maxted}, {Anderson}, {Doyle}, {Gillon},
  {Harrington}, {Iro}, {Jehin}, {Lafreni{\`e}re}, {Smalley}, \&
  {Southworth}}]{Maxted2013}
{Maxted}, P.~F.~L., {Anderson}, D.~R., {Doyle}, A.~P., {et~al.} 2013, \mnras,
  428, 2645

\bibitem[{{Mayne} {et~al.}(2014{\natexlab{a}}){Mayne}, {Baraffe}, {Acreman},
  {Smith}, {Browning}, {Sk{\aa}lid Amundsen}, {Wood}, {Thuburn}, \&
  {Jackson}}]{Mayne2014a}
{Mayne}, N.~J., {Baraffe}, I., {Acreman}, D.~M., {et~al.} 2014{\natexlab{a}},
  \aap, 561, A1

\bibitem[{{Mayne} {et~al.}(2014{\natexlab{b}}){Mayne}, {Baraffe}, {Acreman},
  {Smith}, {Wood}, {Amundsen}, {Thuburn}, \& {Jackson}}]{Mayne2014b}
{Mayne}, N.~J., {Baraffe}, I., {Acreman}, D.~M., {et~al.} 2014{\natexlab{b}},
  Geoscientific Model Development, 7, 3059

\bibitem[{\mbox{Seager}(2010)}]{Seager2010b}
\mbox{Seager}, S. 2010, {Exoplanet Atmospheres: Physical Processes} ({Princeton
  University Press})

\bibitem[{{McCullough} {et~al.}(2014){McCullough}, {Crouzet}, {Deming}, \&
  {Madhusudhan}}]{McCullough2014}
{McCullough}, P.~R., {Crouzet}, N., {Deming}, D., \& {Madhusudhan}, N. 2014,
  \apj, 791, 55

\bibitem[{{Menou} \& {Rauscher}(2009)}]{Menou2009}
{Menou}, K. \& {Rauscher}, E. 2009, \apj, 700, 887

\bibitem[{{Moses} {et~al.}(2011){Moses}, {Visscher}, {Fortney}, {Showman},
  {Lewis}, {Griffith}, {Klippenstein}, {Shabram}, {Friedson}, {Marley}, \&
  {Freedman}}]{Moses2011}
{Moses}, J.~I., {Visscher}, C., {Fortney}, J.~J., {et~al.} 2011, \apj, 737, 15

\bibitem[{{Oreopoulos} {et~al.}(2012){Oreopoulos}, {Mlawer}, {Delamere},
  {Shippert}, {Cole}, {Fomin}, {Iacono}, {Jin}, {Li}, {Manners},
  {R{\"a}Is{\"a}Nen}, {Rose}, {Zhang}, {Wilson}, \& {Rossow}}]{Oreopoulos2012}
{Oreopoulos}, L., {Mlawer}, E., {Delamere}, J., {et~al.} 2012, Journal of
  Geophysical Research (Atmospheres), 117, 6118

\bibitem[{{Parmentier} {et~al.}(2016){Parmentier}, {Fortney}, {Showman},
  {Morley}, \& {Marley}}]{Parmentier2016}
{Parmentier}, V., {Fortney}, J.~J., {Showman}, A.~P., {Morley}, C.~V., \&
  {Marley}, M.~S. 2016, ArXiv e-prints

\bibitem[{{Parmentier} {et~al.}(2013){Parmentier}, {Showman}, \&
  {Lian}}]{Parmentier2013}
{Parmentier}, V., {Showman}, A.~P., \& {Lian}, Y. 2013, \aap, 558, A91

\bibitem[{{Penndorf}(1957)}]{Penndorf1957}
{Penndorf}, R. 1957, Journal of the Optical Society of America (1917-1983), 47,
  176

\bibitem[{{Polichtchouk} \& {Cho}(2012)}]{Polichtchouk2012}
{Polichtchouk}, I. \& {Cho}, J.~Y.-K. 2012, \mnras, 424, 1307

\bibitem[{{Rauscher} \& {Menou}(2010)}]{Rauscher2010}
{Rauscher}, E. \& {Menou}, K. 2010, \apj, 714, 1334

\bibitem[{{Rauscher} \& {Menou}(2012)}]{Rauscher2012}
{Rauscher}, E. \& {Menou}, K. 2012, \apj, 750, 96

\bibitem[{{Redfield} {et~al.}(2008){Redfield}, {Endl}, {Cochran}, \&
  {Koesterke}}]{Redfield2008}
{Redfield}, S., {Endl}, M., {Cochran}, W.~D., \& {Koesterke}, L. 2008, \apjl,
  673, L87

\bibitem[{{Schwarz} {et~al.}(2015){Schwarz}, {Brogi}, {de Kok}, {Birkby}, \&
  {Snellen}}]{Schwarz2015}
{Schwarz}, H., {Brogi}, M., {de Kok}, R., {Birkby}, J., \& {Snellen}, I. 2015,
  ArXiv e-prints

\bibitem[{{Showman} {et~al.}(2008){Showman}, {Cooper}, {Fortney}, \&
  {Marley}}]{Showman2008}
{Showman}, A.~P., {Cooper}, C.~S., {Fortney}, J.~J., \& {Marley}, M.~S. 2008,
  \apj, 682, 559

\bibitem[{{Showman} {et~al.}(2009){Showman}, {Fortney}, {Lian}, {Marley},
  {Freedman}, {Knutson}, \& {Charbonneau}}]{Showman2009}
{Showman}, A.~P., {Fortney}, J.~J., {Lian}, Y., {et~al.} 2009, \apj, 699, 564

\bibitem[{{Showman} \& {Guillot}(2002)}]{Showman2002}
{Showman}, A.~P. \& {Guillot}, T. 2002, \aap, 385, 166

\bibitem[{{Showman} \& {Polvani}(2011)}]{Showman2011b}
{Showman}, A.~P. \& {Polvani}, L.~M. 2011, \apj, 738, 71

\bibitem[{{Sing} {et~al.}(2011){Sing}, {D{\'e}sert}, {Fortney}, {Lecavelier Des
  Etangs}, {Ballester}, {Cepa}, {Ehrenreich}, {L{\'o}pez-Morales}, {Pont},
  {Shabram}, \& {Vidal-Madjar}}]{Sing2011}
{Sing}, D.~K., {D{\'e}sert}, J.-M., {Fortney}, J.~J., {et~al.} 2011, \aap, 527,
  A73

\bibitem[{{Sing} {et~al.}(2016){Sing}, {Fortney}, {Nikolov}, {Wakeford},
  {Kataria}, {Evans}, {Aigrain}, {Ballester}, {Burrows}, {Deming},
  {D{\'e}sert}, {Gibson}, {Henry}, {Huitson}, {Knutson}, {Etangs}, {Pont},
  {Showman}, {Vidal-Madjar}, {Williamson}, \& {Wilson}}]{Sing2016}
{Sing}, D.~K., {Fortney}, J.~J., {Nikolov}, N., {et~al.} 2016, \nat, 529, 59

\bibitem[{{Sing} {et~al.}(2012){Sing}, {Huitson}, {Lopez-Morales}, {Pont},
  {D{\'e}sert}, {Ehrenreich}, {Wilson}, {Ballester}, {Fortney}, {Lecavelier des
  Etangs}, \& {Vidal-Madjar}}]{Sing2012}
{Sing}, D.~K., {Huitson}, C.~M., {Lopez-Morales}, M., {et~al.} 2012, \mnras,
  426, 1663

\bibitem[{{Sing} {et~al.}(2015){Sing}, {Wakeford}, {Showman}, {Nikolov},
  {Fortney}, {Burrows}, {Ballester}, {Deming}, {Aigrain}, {D{\'e}sert},
  {Gibson}, {Henry}, {Knutson}, {Lecavelier des Etangs}, {Pont},
  {Vidal-Madjar}, {Williamson}, \& {Wilson}}]{Sing2015}
{Sing}, D.~K., {Wakeford}, H.~R., {Showman}, A.~P., {et~al.} 2015, \mnras, 446,
  2428

\bibitem[{{Snellen} {et~al.}(2008){Snellen}, {Albrecht}, {de Mooij}, \& {Le
  Poole}}]{Snellen2008}
{Snellen}, I.~A.~G., {Albrecht}, S., {de Mooij}, E.~J.~W., \& {Le Poole}, R.~S.
  2008, \aap, 487, 357

\bibitem[{{Snellen} {et~al.}(2010){Snellen}, {de Kok}, {de Mooij}, \&
  {Albrecht}}]{Snellen2010}
{Snellen}, I.~A.~G., {de Kok}, R.~J., {de Mooij}, E.~J.~W., \& {Albrecht}, S.
  2010, \nat, 465, 1049

\bibitem[{{Spiegel} {et~al.}(2009){Spiegel}, {Silverio}, \&
  {Burrows}}]{Spiegel2009}
{Spiegel}, D.~S., {Silverio}, K., \& {Burrows}, A. 2009, \apj, 699, 1487

\bibitem[{{Stevenson} {et~al.}(2014){Stevenson}, {D{\'e}sert}, {Line}, {Bean},
  {Fortney}, {Showman}, {Kataria}, {Kreidberg}, {McCullough}, {Henry},
  {Charbonneau}, {Burrows}, {Seager}, {Madhusudhan}, {Williamson}, \&
  {Homeier}}]{Stevenson2014}
{Stevenson}, K.~B., {D{\'e}sert}, J.-M., {Line}, M.~R., {et~al.} 2014, Science,
  346, 838

\bibitem[{{Swain} {et~al.}(2008){Swain}, {Bouwman}, {Akeson}, {Lawler}, \&
  {Beichman}}]{Swain2008}
{Swain}, M.~R., {Bouwman}, J., {Akeson}, R.~L., {Lawler}, S., \& {Beichman},
  C.~A. 2008, \apj, 674, 482

\bibitem[{{Tennyson} \& {Yurchenko}(2012)}]{Tennyson2012}
{Tennyson}, J. \& {Yurchenko}, S.~N. 2012, \mnras, 425, 21

\bibitem[{{Tennyson} {et~al.}(2016){Tennyson}, {Yurchenko}, {Al-Refaie},
  {Barton}, {Chubb}, {Coles}, {Diamantopoulou}, {Gorman}, {Hill}, {Lam},
  {Lodi}, {McKemmish}, {Na}, {Owens}, {Polyansky}, {Sousa-Silva}, {Underwood},
  {Yachmenev}, \& {Zak}}]{Tennyson2016}
{Tennyson}, J., {Yurchenko}, S.~N., {Al-Refaie}, A.~F., {et~al.} 2016, Journal
  of Molecular Spectroscopy

\bibitem[{{Thomas} \& {Stamnes}(2002)}]{Thomas2002}
{Thomas}, G.~E. \& {Stamnes}, K. 2002, {Radiative Transfer in the Atmosphere
  and Ocean} ({Cambridge University Press})

\bibitem[{{Thrastarson} \& {Cho}(2010)}]{Thrastarson2010}
{Thrastarson}, H.~T. \& {Cho}, J.~Y. 2010, \apj, 716, 144

\bibitem[{{Toon} {et~al.}(1989){Toon}, {McKay}, {Ackerman}, \&
  {Santhanam}}]{Toon1989}
{Toon}, O.~B., {McKay}, C.~P., {Ackerman}, T.~P., \& {Santhanam}, K. 1989,
  \jgr, 94, 16287

\bibitem[{{Tremblin} {et~al.}(2016){Tremblin}, {Amundsen}, {Chabrier},
  {Baraffe}, {Drummond}, {Hinkley}, {Mourier}, \& {Venot}}]{Tremblin2016}
{Tremblin}, P., {Amundsen}, D.~S., {Chabrier}, G., {et~al.} 2016, \apjl, 817,
  L19

\bibitem[{{Tremblin} {et~al.}(2015){Tremblin}, {Amundsen}, {Mourier},
  {Baraffe}, {Chabrier}, {Drummond}, {Homeier}, \& {Venot}}]{Tremblin2015}
{Tremblin}, P., {Amundsen}, D.~S., {Mourier}, P., {et~al.} 2015, \apjl, 804,
  L17

\bibitem[{{Wakeford} {et~al.}(2013){Wakeford}, {Sing}, {Deming}, {Gibson},
  {Fortney}, {Burrows}, {Ballester}, {Nikolov}, {Aigrain}, {Henry}, {Knutson},
  {Lecavelier des Etangs}, {Pont}, {Showman}, {Vidal-Madjar}, \&
  {Zahnle}}]{Wakeford2013}
{Wakeford}, H.~R., {Sing}, D.~K., {Deming}, D., {et~al.} 2013, \mnras, 435,
  3481

\bibitem[{{Wong} {et~al.}(2015){Wong}, {Knutson}, {Lewis}, {Kataria},
  {Burrows}, {Fortney}, {Schwartz}, {Agol}, {Cowan}, {Deming}, {D{\'e}sert},
  {Fulton}, {Howard}, {Langton}, {Laughlin}, {Showman}, \&
  {Todorov}}]{Wong2015}
{Wong}, I., {Knutson}, H.~A., {Lewis}, N.~K., {et~al.} 2015, ArXiv e-prints

\bibitem[{{Wood} {et~al.}(2014){Wood}, {Staniforth}, {White}, {Allen},
  {Diamantakis}, {Gross}, {Melvin}, {Smith}, {Vosper}, {Zerroukat}, \&
  {Thuburn}}]{Wood2014}
{Wood}, N., {Staniforth}, A., {White}, A., {et~al.} 2014, Q. J. R. Meteorol.
  Soc., 140, 1505

\bibitem[{{Yurchenko} \& {Tennyson}(2014)}]{Yurchenko2014}
{Yurchenko}, S.~N. \& {Tennyson}, J. 2014, \mnras, 440, 1649

\bibitem[{{Zdunkowski} \& {Korb}(1985)}]{Zdunkowski1985}
{Zdunkowski}, W.~G. \& {Korb}, G.~J. 1985, Promet, 2/3, 26

\bibitem[{{Zdunkowski} {et~al.}(1980){Zdunkowski}, {Welch}, \&
  {Korb}}]{Zdunkowski1980}
{Zdunkowski}, W.~G., {Welch}, R.~M., \& {Korb}, G. 1980, Beitr\"{a}ge zur
  Physik der Atmosph\"{a}re, 53, 147

\bibitem[{{Zellem} {et~al.}(2014){Zellem}, {Lewis}, {Knutson}, {Griffith},
  {Showman}, {Fortney}, {Cowan}, {Agol}, {Burrows}, {Charbonneau}, {Deming},
  {Laughlin}, \& {Langton}}]{Zellem2014}
{Zellem}, R.~T., {Lewis}, N.~K., {Knutson}, H.~A., {et~al.} 2014, \apj, 790, 53

\end{thebibliography}

\end{document}